\documentclass[10pt]{article}

\usepackage{a4wide}
\usepackage{authblk}
\usepackage{graphicx}
\usepackage{epstopdf, epsfig}	
\usepackage{amsmath}
\usepackage{amssymb}
\usepackage[round,sort&compress,comma,authoryear]{natbib}
\usepackage{float}
\usepackage{hyperref}
\usepackage{lscape}
\usepackage{xcolor}
\usepackage{newtxtext}
\usepackage{newtxmath}
\usepackage{natbib}
\usepackage{hyperref}
\usepackage{fdsymbol}
\hypersetup{
    colorlinks = true,
    urlcolor   = blue,
    citecolor  = black,
}

\newcommand{\RomanNumeralCaps}[1]
\linenumbers

\title{Ribbing patterns in inertial rotary drag-out}

\author[1, $\dagger$]{{J. John Soundar Jerome}}
\author[1]{Pierre Trontin}
\author[1]{Jean-Philippe Matas} 
\affil[1]{Universit\'{e} de Lyon, Universit\'{e} Claude Bernard Lyon $1$, Laboratoire de M\'{e}canique des Fluides et d'Acoustique, CNRS UMR--$5509$, Boulevard $11$ novembre, F--$69622$ Villeurbanne cedex, LYON, France}
\affil[$\dagger$]{email : john-soundar@univ-lyon1.fr}

\date{}
\begin{document}

\maketitle

\begin{abstract}
We report pattern formation in an otherwise non-uniform and unsteady flow arising in high-speed liquid entrainment conditions on the outer wall of a wide rotating drum. We show that the coating flow in this rotary drag-out undergoes axial modulations to form an array of roughly vertical thin liquid sheets which slowly drift from the middle of the drum towards its side walls. Thus, the number of sheets fluctuate in time such that the most probable rib spacing varies ever so slightly with the speed, and a little less weakly with the viscosity. We propose that these axial patterns are generated due to a primary instability driven by an adverse pressure gradient in the meniscus region of the rotary drag-out flow, similar to the directional Saffman-Taylor instability, as is well-known for ribbing in film-splitting flows. {Rib spacing based on this mechanistic model turns out to be proportional to the capillary length, wherein the scaling factor can be determined based on existing models for film entrainment at both low and large Capillary numbers.} In addition, we performed Direct Numerical Simulations, which reproduce the experimental phenomenology and the associated wavelength. We further include two numerical cases wherein either the liquid density, or the liquid surface tension are quadrupled while keeping all other parameters identical with experiments. The rib spacings of these cases are in agreement with the predictions of our model.
\end{abstract}

\section{Introduction}
\label{sec:intro}
A gas/liquid interface is rarely uniform. It is often {irregular} with defects such as bubbles and drops. It is nevertheless common that two-phase flows, even in the most complex situations, exhibit robust patterns such as waves, rivulets, cells, ridges, and finger-like structures. In many cases, the primary characteristics of such pattern formation can be elucidated through the analysis of the flow stability, and/or via phenomenological models \citep{CrossHohenberg_PatternFormation_RevModPhy1993, Fauve_PatternFormingInsta_1998, GallaireBrun_Patterns_2017}. An example of ordered pattern formation in an otherwise non-uniform and seemingly random free-surface flow is reported here, for the case of liquid film entrainment along a rotating wheel. {This is also relevant to understand the liquid load that be can left over a moving substrate during drag-out, as for example, by car wheels on a wet road, journal bearings at high-operational speeds, and rapid roll coating processes which exhibit \textit{misting} \i.e., atomization of the entrained liquid film flow.}

\begin{figure}
\begin{center}
\includegraphics[width=1\textwidth,keepaspectratio=true]{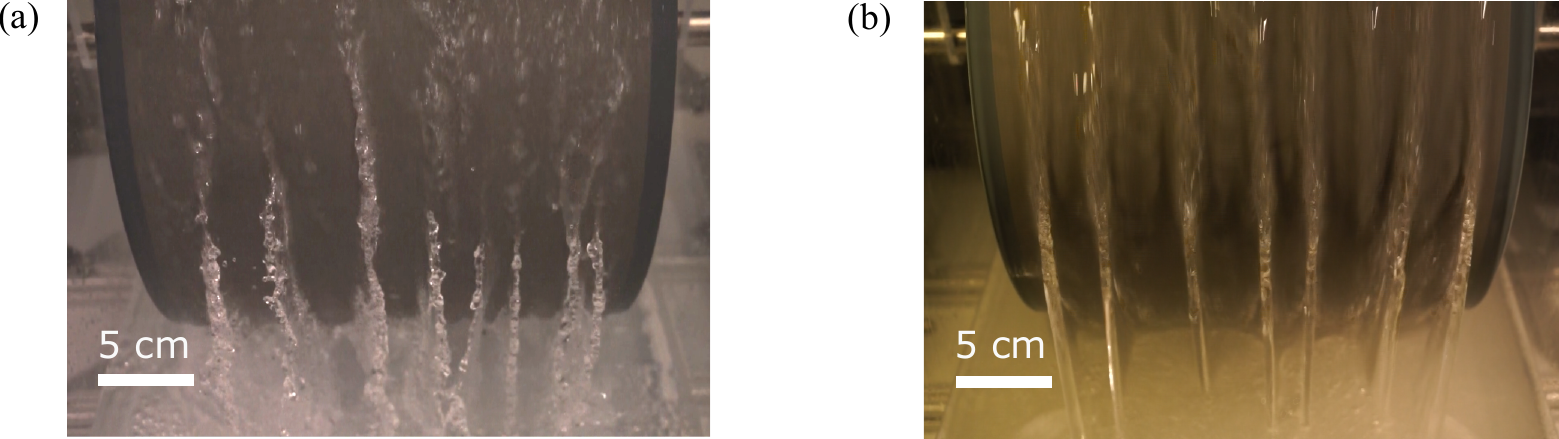}
\end{center}
\caption{Examples of ribbing patterns \i.e., sheets of liquid rise vertically out of the reservoir in relatively regular spacing along the axis of a rotating drum in (a) water, linear velocity $3.8$~m/s and (b) water/UCON mixture of approximately $100$ times the viscosity of water, $1.6$~m/s. Both photos correspond to the case when the working liquid covers $10$\% of the drum radius at rest.}
\label{fig:visu_nappes}
\end{figure}

\begin{figure}[!h]
	\begin{center}
		\includegraphics[width=1\textwidth,keepaspectratio=true]{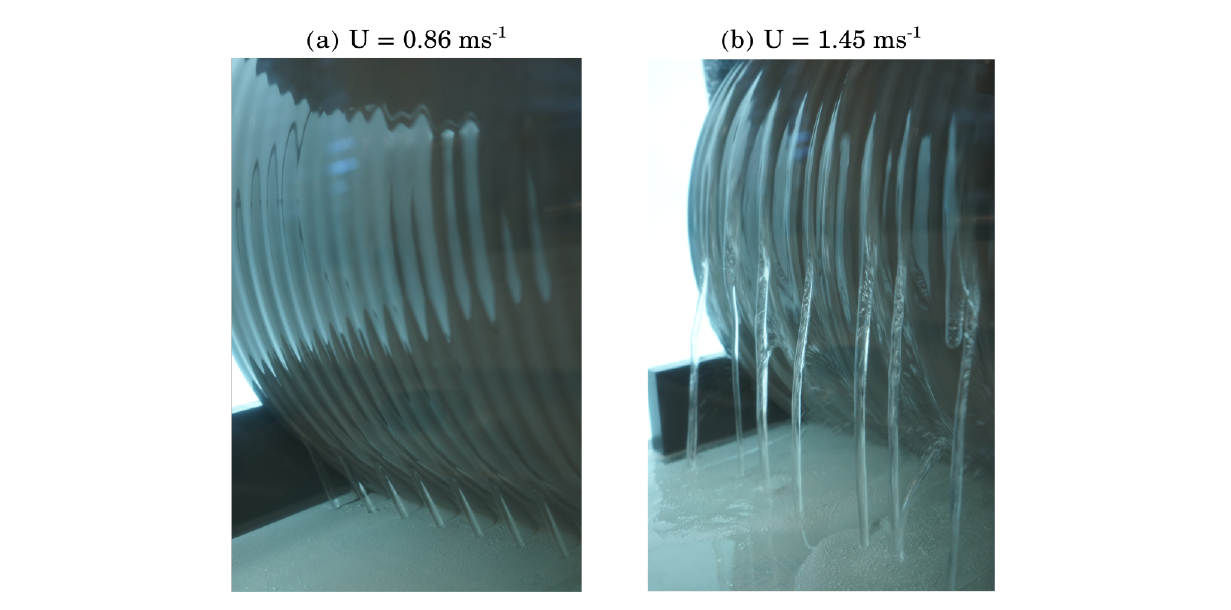}
	\end{center}
\caption{{Images of ribbing patterns at two different speeds in the viscous UCON/water mixture \textit{(Photo credits: Rosie Cates)} for the same wheel immersion conditions as above.}}
\label{fig:ribs_persp}
\end{figure}

We consider a circular drum of radius $R$ which is only partly immersed in a liquid of viscosity $\mu$, density $\rho$ and surface tension $\sigma$. Let $\Omega$ be its angular speed. Analogous to the celebrated $2$D plate drag-out problem treated by \cite{levich1942dragging, derjagin1943thickness}, {the rotating drum} entrains a liquid film over its outer surface if its linear speed $U = R \Omega$ is more than the dewetting speed. Recently, we pointed out that as the drum is rotated at faster rates one or more liquid sheets appear on the emerging side depending on the drum width \citep[see Fig. 4 \& 5, and references therein]{jerome2021inertial}. Here in Fig. \ref{fig:visu_nappes}, we display instantaneous images of an array of liquid sheets that occur for a much wider drum in our present experiments, and for two liquids which differ in their dynamic viscosity by a factor $100$. {As the drum speed is increased steadily, the quasi-static meniscus along the axis of the cylinder deforms into a regular pattern of thin liquid sheets with a thick rim (see Fig. \ref{fig:ribs_persp}). The sheet height increases as the square of the drum speed, similar to the case of a single sheet in front of a horizontal rotating disc in a liquid bath \citep{jerome2021inertial}.} The goal of the present work is to elucidate the mechanism at the origin of these liquid sheets. 


\citet{yih1960instability} was the first to consider the theoretical prediction of ribbing patterns in the context of {Fourdrinier machines}\footnote{{A device that features a rotating drum in a tank of pulp and water for producing a continuous web of paper by draining the mixture on a specially meshed conveyor belt.}} in the paper industry. He performed a linear stability analysis of a base flow which consists of a thin, freely-rotating film in the absence of gravity and viscous shear {at the liquid/gas interface}. Thereby, he observed that the critical wavelength $\lambda_Y$ is only weakly, or negligibly dependent on the film Reynolds number $Re_f = (UR/\nu) (\delta_f/R)^2$ where $\delta_f$ is the film thickness which was manually metered {in} his experiments. From both theory and experiments, \citet{yih1960instability} further demonstrated that $\lambda_Y$ is strongly dependent on wheel radius, angular speed and the liquid surface tension $\sigma$ via the Weber number $We = \rho U^2 R/\sigma$ so that $\lambda_Y \simeq 2 \pi R \sqrt{3/We}$. This is primarily due to the fact that the driving mechanism for the instability is the centripetal acceleration. So, Yih's ribbing instability is analogous to Taylor vortices \citep{Taylor1923} occurring in a liquid bound between two rotating cylinders, wherein the outer cylinder wall is replaced by a free surface boundary condition. 
%Also, note that Yih's critical wavelength can be rewritten as $\lambda_Y \simeq 2 \pi l_c \sqrt{3g/\left(R\Omega^2\right)}$ where $l_c = \sqrt{\sigma/(\rho g)}$ is the capillary length. 
{A similar axial instability is also known in the case of \textit{external} rimming flows \cite[see Fig. 5--9]{Moffatt1977}. While analyzing the existence of steady solutions for the case of a slowly-rotating thin viscous film on the outer surface of a circular cylinder, \cite{Moffatt1977} observed \textit{ring-like} patterns along the axis of the cylinder in his experiments. Surprisingly, these patterns occurred even at speeds as small as $12$~rpm. He suggested that such patterns in a viscous rotating film are not only due to centrifugal forces but also driven by gravity and shear forces, which are neglected in \cite{yih1960instability}'s analysis. To the authors' present knowledge, the role of centrifugal forces in such a simple wall-driven viscous film flow is still an open question.}
\begin{figure}
\begin{center}
\includegraphics[width=1\textwidth,keepaspectratio=true]{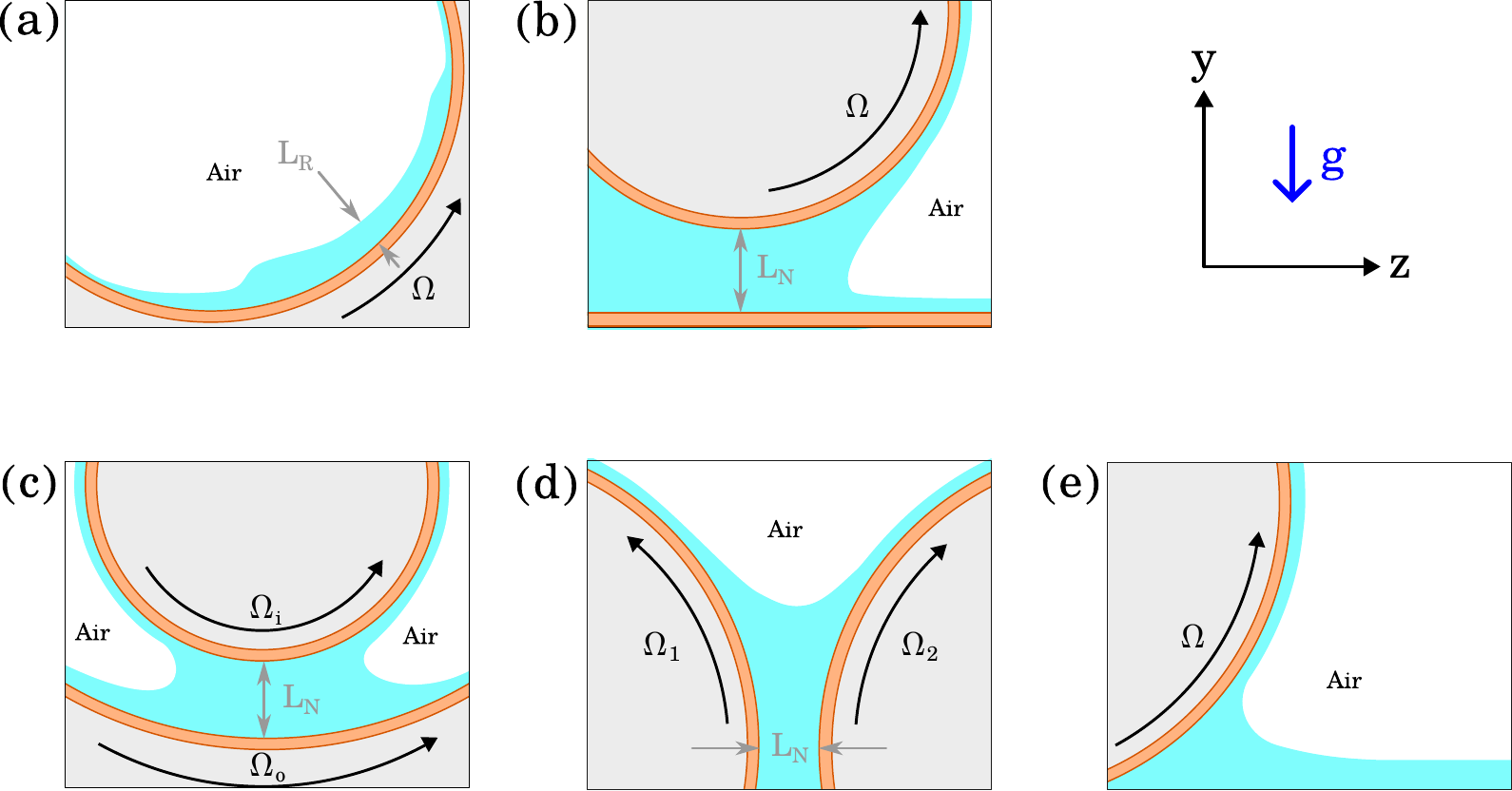}
\end{center}
\caption{Schematic view (not to scale) of previous investigations on axial pattern formation in coating flows on the outside and the inside of a cylinder for (a) \textit{rimming flows} with liquid ridge as elucidated by \citet{Hosoi_RibRimming_1999PoF}, (b) -- (d) \textit{film-splitting flows} resulting in a thin film on one or more a rotating substrate as studied by \citet{adachi1988coating}, \citet{rabaud1994dynamiques}, \citet{coyle1990stability}, respectively. (e) The present configuration is analogous to the film-splitting case but the meniscus region is free to evolve away from the cylinder. Also, the Reynolds number based on the film thickness is about $10^2$--$10^4$ in our study.}
\label{fig:coatingflows}
\end{figure}
{Many investigations of ribbing and cellular patterns can be found in the case of \textit{internal} rimming flows \citep{Balmer_Hygrocyst_Nature1970, ThoroddsenMahadevan_ExpFluids1997} wherein a fixed quantity of liquid inside a cylinder is subject to rotation, see Fig. \ref{fig:coatingflows}(a).} \citet{Hosoi_RibRimming_1999PoF} analysed the rimming flow case for small volume fraction of liquid i.e., up to $6$\% of the cylinder volume. By including inertia, along with gravity, capillarity and viscous forces, they attributed the onset of the axial instability to the presence of a localized liquid ridge {of reverse flow} in the thin film, as depicted in Fig. \ref{fig:coatingflows}(a). Such a ridge appears when the rescaled Reynolds number $Re = \epsilon U R/\nu$ is of $\mathcal{O}(1)$ where $\nu = \mu/\rho$ is the liquid kinematic viscosity and $\epsilon = \delta_g/R$ is a small parameter, if $\delta_g = \sqrt{\nu U/g}$ is the relevant length scale for the film thickness due to a balance between gravity and viscous shear arising from the moving wall. {Thus, any perturbation in the axial direction leads to thick film regions draining much faster into the ridge than thinner ones, since gravity effects are stronger in the former while the wall shear stress is more important in the latter region.} %Thus, any small perturbation in the axial direction leads to thick regions of liquid falling down much faster than thinner ones due to stronger shear in the latter.
Thereby, the ridge height continues to vary along the cylinder axis until surface tension forces become significant due to the free-surface curvature along the cylinder axis. In addition, the authors proposed that the wavelength of the most unstable mode is proportional to $L_R Ca^{-1/3}$, where $Ca = \mu U / \sigma$ is the capillary number and $L_R$ is some typical ridge height. 
%For the case of water in Fig. \ref{fig:visu_nappes}(a) and water/UCON mixture in Fig. \ref{fig:visu_nappes}(b), the capillary number is $\mathcal{O}(10^{-2})$ and $\mathcal{O}(1)$, respectively. If $L_R$ is taken to be of the order of $\delta_g$, then it is expected that the distance between the sheets will increase as $(\mu U)^{1/6} (\rho g)^{1/2} \sigma^{1/3} $ i.e., by a factor of approximately two for the viscous water/UCON mixture compared to water. However, the photos in Fig. \ref{fig:visu_nappes} show that the number of sheets is almost unchanged for these two cases. {the model of figure 10 below predicts the same factor 2: are they related?}

Ribbing was also often observed in film-splitting flows \citep{schweizer2012liquid} that occur when the outer wall of one or more rollers drag a liquid film from a reservoir, in the presence of a strong confinement. A few such thin viscous coating flow configurations are illustrated in Fig. \ref{fig:coatingflows}(b)--(d), where the {minimum gap is usually much smaller than the roller radius \i.e., $L_N \ll R$}. {In this context, a large-body of investigation is available for what is now commonly referred to as the ``printer's instability'', wherein a regular axial variation of film thickness occurs in the widening gap between two rotating cylinders, Fig. \ref{fig:coatingflows}(c) \citep{pitts1961flow, mill1967formation, Savage1977, Hakim1990, coyle1990stability, rabaud1994dynamiques}.} Ribbing in this configuration  is often discerned when the capillary number is sufficiently large, depending on the confinement ratio $L_N/R$. For a given liquid viscosity and surface tension, the number of ribs ($n_r$) increases rapidly with the roller speed and then it saturates at larger velocities. Also, a smaller number of ribs are observed as rollers are kept further apart i.e., $L_N/R$ is increased. The {gap} $L_N$ is typically of the order of 10~$\mu$m to $1$~mm, and the distance between ribs varies between a few hundreds of micrometers to a few millimetres. Gravity can be disregarded in the nip and a Couette-Poiseuille flow is imposed in the neighbourhood of the meniscus, resulting from a balance between the wall-imposed shear stress and {mass conservation}. 
%Since the curvature of the free-surface changes signs as we move from the meniscus zone to the thin film zone over the rollers, 
An adverse pressure gradient is therefore present close to the interface, and leads to the formation of fingers which progressively stretch away from the meniscus. {Ribbing in this configuration is analogous to the directional Saffman-Taylor instability \citep{SaffmanTaylor1958, Hakim1990, rabaud1994dynamiques, Reinelt_DirectionalViscFingering_JFM1995}}.
%{where the air/liquid interface is stationary. 
In general, the rib spacing $\lambda$ decreases as $Ca^{-1/2}$ when the capillary number is increased. But at sufficiently large capillary numbers, it saturates to a constant value depending on the {gap} as $\lambda \propto L_N^{2/3}$, where the constant of proportionality is a function of the liquid physical properties, and the nip geometry \citep[see Fig. 6 \& 7]{adachi1988coating}, since the latter precisely fixes both the meniscus location and the adverse pressure gradient experienced by the static air/liquid interface. {Finally, these regular ribbing patterns become unstable and result in spatio-temporal chaos as the capillary number is further increased \citep{Couder1990, Michalland1996}.}
%ADD A SENTENCE REGARDING IMPACT of SIGMA/LOCATION OF MENISCUS%
%any axial modulation in film thickness, or the meniscus position, leads to a decrease (increase) in the Laplace pressure in thinner (thicker) film-splitting regions. Thereby, a higher adverse pressure gradient occurs in places where the rising film is thin, as compared to that in the zones where it is thicker. For a given wall shear stress, film thickness should then diminish around troughs, while it should grow near the crests of such meniscus perturbations if liquid surface tension could not rapidly alleviate such interface modulations. This mechanism was first put forward by \citet{pitts1961flow}. 
%This interfacial instability due to an adverse pressure gradient in the lubrication flow neighbourhood of a stationary air/liquid interface can be interpreted as an instability of an interface between two immiscible fluids whose viscosity decreases in the direction parallel to the one where the pressure gradient is favourable. In these terms, ribbing is analogous to directional Saffman-Taylor instability \cite{SaffmanTaylor1958}. 

{A related pattern formation occurs in liquid film flows over the end of a flat or a curved plate. Depending on the mass flux imposed by the source, \cite{Pritchard1986} illustrated that the overflowing liquid film flow could break into dripping droplets and liquid columns along the streamwise direction.  At larger flow rates, a rich dynamical behaviour is observed \citep[see references therein]{Giorgiutti1995, Limat1997}. The underlying mechanism is due to the Rayleigh-Taylor instability when a thin liquid film is accelerated towards a less denser fluid. This is also at the origin of patterns known as rivulets on thin viscous films flowing under an inclined flat or a curved plate \citep{Balestra2018, Ledda2020, Lerisson2020}. In all these cases, the characteristic wavelength of the primary instability is proportional to the film thickness which is fixed by the source which feeds the film flow, or the initial coating film thickness. Furthermore, the wall is static whereby the film flow is driven only by gravity in the absence of an externally imposed wall shear.}

The present article concerns ribbing patterns such as those shown in Fig. \ref{fig:visu_nappes} that arise from a situation of coating flows depicted in Fig. \ref{fig:coatingflows}(e). {The aforementioned draining, rimming and film-splitting flows belong to a class of coating flows where the Reynolds lubrication approximation is valid \i.e., the Reynolds number $Re_f = (U R / \nu) (\delta_f /R)^2$ which is based on the linear velocity $U$ and the entrained liquid film thickness $\delta_f$ is either small or $\mathcal{O}(1)$ at best. Inertia plays little role in these wall-driven liquid entrainment flows, except maybe for the ridge formation in internal rimming flows as in Fig \ref{fig:coatingflows}(a). In our study, we are interested in flow regimes where inertia is no longer negligible since $Re_f$ becomes as large as $10^4$ in some cases. Fig. \ref{fig:visu_nappes}(a) shows that the rims of the liquid sheets break-up and result in atomization. So, in many experiments in water and at high-speed cases in the viscous water/UCON mixture, droplets detach from ribs, and air bubbles which are entrained in the plunging film below the wheel are re-entrained within the ribs. Also, the ribs themselves experience strong fluctuations in shape and position.}

{Under these conditions, full-scale numerical simulations are difficult and costly. So, a phenomenological approach is adopted here to investigate pattern formation in liquid drag-out flow at very large Reynolds numbers ($Re_f \gg 1$).}
%Our previous work \citep{jerome2021inertial} suggests that the \textit{time-averaged} mass flux in high-speed liquid entrainment flows can be estimated via an analogy with the classical $2$D coating film flow rate based on Landau-Levich-Deryaguin vertical plate drag-out problem.  So, as in \cite{jerome2021inertial}, } Our study could be of some importance not only to understand the origin of secondary flow structures developed during inertial drag-out but also to estimate the maximum liquid load that can remain on a moving substrate during drag-out, as for example, by car wheels on a wet road, journal bearings at high-operational speeds, and rapid roll coating process which exhibit \textit{misting} \i.e., atomization of the entrained liquid film flow.
We begin by presenting our experimental set-up and, then by describing our observations and results in section \ref{sec:exp_obs}. {In section \ref{sec:STmechanism}, we propose a mechanism based on the \textit{directional} Saffman-Taylor instability in order to predict the characteristic distance between ribs.} Finally, we present in section \ref{sec:numMethods} the results of a numerical simulation of this problem, and discuss the consistency of the obtained results  with our model.

\section{Experimental observations}
\label{sec:exp_obs}
\subsection{Materials, Set-up and Methods}
\label{sec:set-up}
\begin{figure}[!h]
\begin{center}
\includegraphics[width=1\textwidth,keepaspectratio=true]{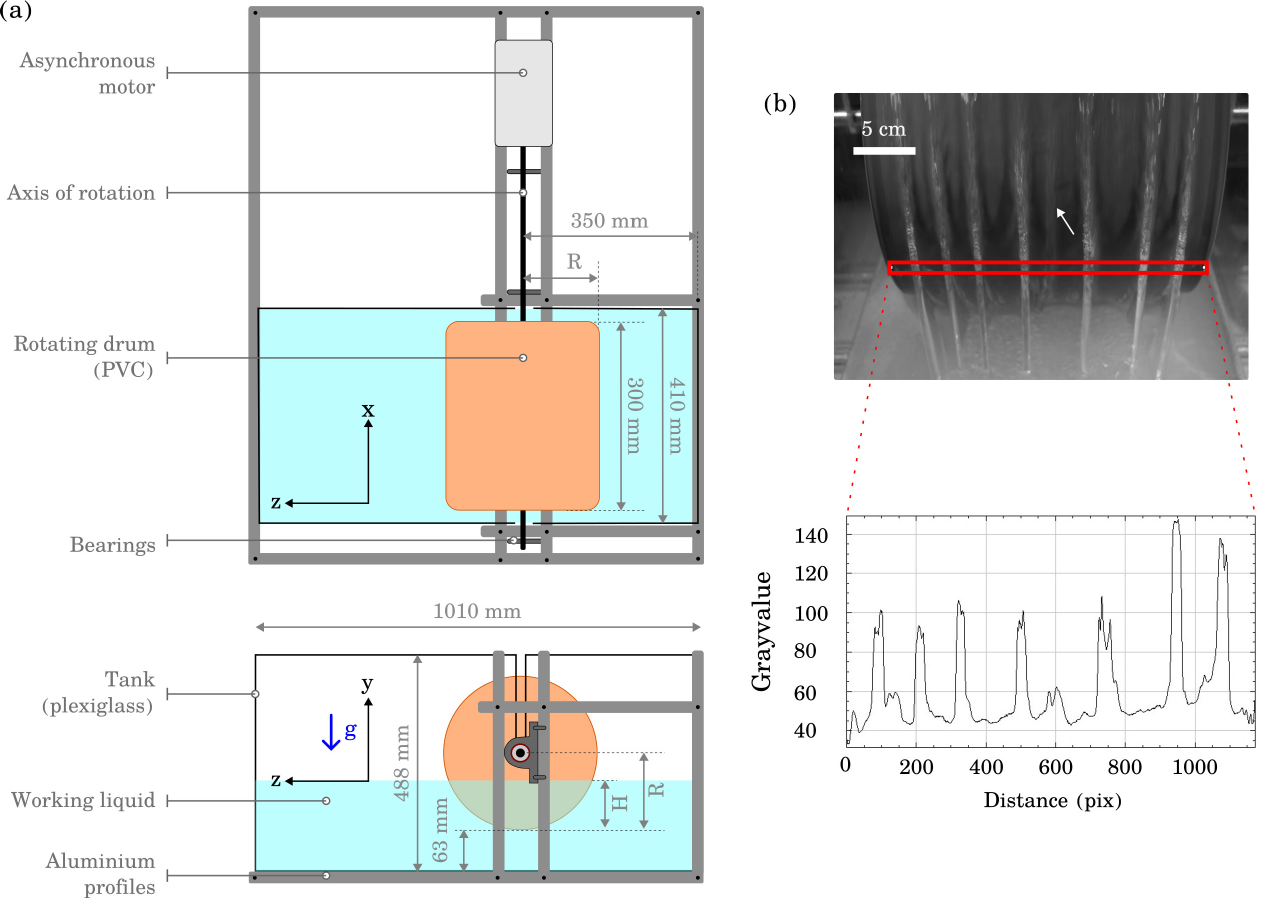}
\end{center}
\caption{(a) Top and side view schematics of experimental setup depicting the liquid tank and the partially-immersed rotating drum. Water depth can be changed to control the drum immersion, such that $H/R \in [0, 1]$. (b) Snapshot displaying the front view of a typical ribbing pattern and the corresponding intensity profile for the case of water/UCON mixture when $H/R = 0.1$ and $U = 1.6$~m/s {(see Supplemental video I, II \& III)}. The white arrow indicates an incipient rib between two other ribs which are drifting apart from each other.}
\label{fig:Setup}
\end{figure}

The set-up is illustrated in Fig. \ref{fig:Setup}. Here, a wheel of radius $R = 19.5$~cm and width $30$~cm, partially immersed in a reservoir, is rotated about its horizontal axis, entrained by an asynchronous motor, coupled with a Parker AC10 AC inverter drive. {The wheel is made out of a thick PVC tube and its axis is a rigid steel bar of $25$~mm diameter which is one meter long. It is mounted on three ball bearings, with a pair of them on each side of the drum (approx. $40$~cm) and the last one closer to the motor. We regularly verified that no visible precession about the cylinder axis occurred when the drum is under rotation.} {In the following, the distance between the wheel bottom and the free surface of the liquid at rest is denoted by $H$.} {The bottom of the wheel is located at $6.3$~cm above the tank floor.} {For all our experiments, we use either water, or a mixture of water and UCON oil (300g of UCON for 700g of water).} The composition of the latter is controlled before {each series of experiments by properly measuring} the mixture density with an Anton Paar DMA35 densimeter (precision $10^{-4}$~g/cm$^3$), and the temperature of the mixture.  {Note that the dynamic viscosity ($\mu$) of the water/UCON mixture can be as high as $100$ times that of water (see Table \ref{tab:LiquidProperties}).} But variations in temperature between different series of experiments with the water/UCON mixture can lead to changes in viscosity of the order of $15$~\%. {So, throughout the article, whenever the experimental results for a water/UCON mixture are presented, the reader shall be referred to its corresponding viscosity value.} {Special care was taken to wash the tank thoroughly every time we changed the liquid from the water/UCON mixture to just water. Traces of UCON in water readily led to vigorous foaming without any marked influence on the unsteady, non-uniform character of the observed ribbing. The effect of surfactant on the rib shape and sheet break-up is beyond the scope of present study.}

Images of the liquid structures generated when the liquid is dragged out by the wheel are recorded with the help of {a} Sony $\alpha$7 camera. A typical image of ribbing patterns recorded by the camera is displayed in Fig. \ref{fig:Setup}(b)-top. We can extract from such an image the associated intensity profiles along the width of the wheel (Fig. \ref{fig:Setup}(b)-bottom).  {Both the image and its intensity profile} are then analysed with Matlab to identify the location of each rib for each image. {This procedure is carried out for at least $1000$ images for experiments with water/UCON mixtures, {and} $1500$ images for experiments with water. Each of these image counts correspond to a video duration of at least $40$ and $60$ seconds, respectively.}

\begin{table}[!h]
  \begin{center}
\def~{\hphantom{0}}
  \begin{tabular}{cccc}
      Liquid  & Density $(\rho)$ & Viscosity $(\mu)$ &Surface tension $(\sigma)$\\[3pt]
         &kg.m$^{-3}$ &$\times 10^{-3}$ Pa.s   &$\times 10^{-3}$ N.m\\[1pt]
         \hline
       {}   &	{}&	{}&		{}\\
       Water   & $999\pm 1$ &$1.05 \pm 0.02$   &$70\pm 2$   \\
      % {}   &	{}&	{}&		{}\\
       70\% Water-30\% UCON$^{\mbox{{\tiny TM}}}$ $3$ ($WU3$)  	& $1045\pm 1$  &$90 \pm 9$   &$58\pm 2$  \\
       {}   &	{}&	{}&		{}\\
  \end{tabular}
  \caption{Properties of liquids used in the present work. {UCON$^{\mbox{{\tiny TM}}}$ Lubricant $75$-H-$90$,$000$ was used for all mixtures used here.}}
  \label{tab:LiquidProperties}
  \end{center}
\end{table}

\subsection{Evolution of rib spacing over time}
%Assuming a rib thickness of the order of the millimeter, the Reynolds number is expected to be of the order of $10^3$ for the water ribs of figure \ref{fig:visu_nappes}a, around which liquid fragmentation is observed, while it is of order 10 for those of figure \ref{fig:visu_nappes}b. 

%Kinematic waves leading to rivulet-like structures \citep{GallaireBrun_Patterns_2017} appear initially for very small wheel immersion and small drum speeds but they are quickly replaced by ribbing patterns. 

\begin{figure}
\begin{center}
\includegraphics[width=1\textwidth,keepaspectratio=true]{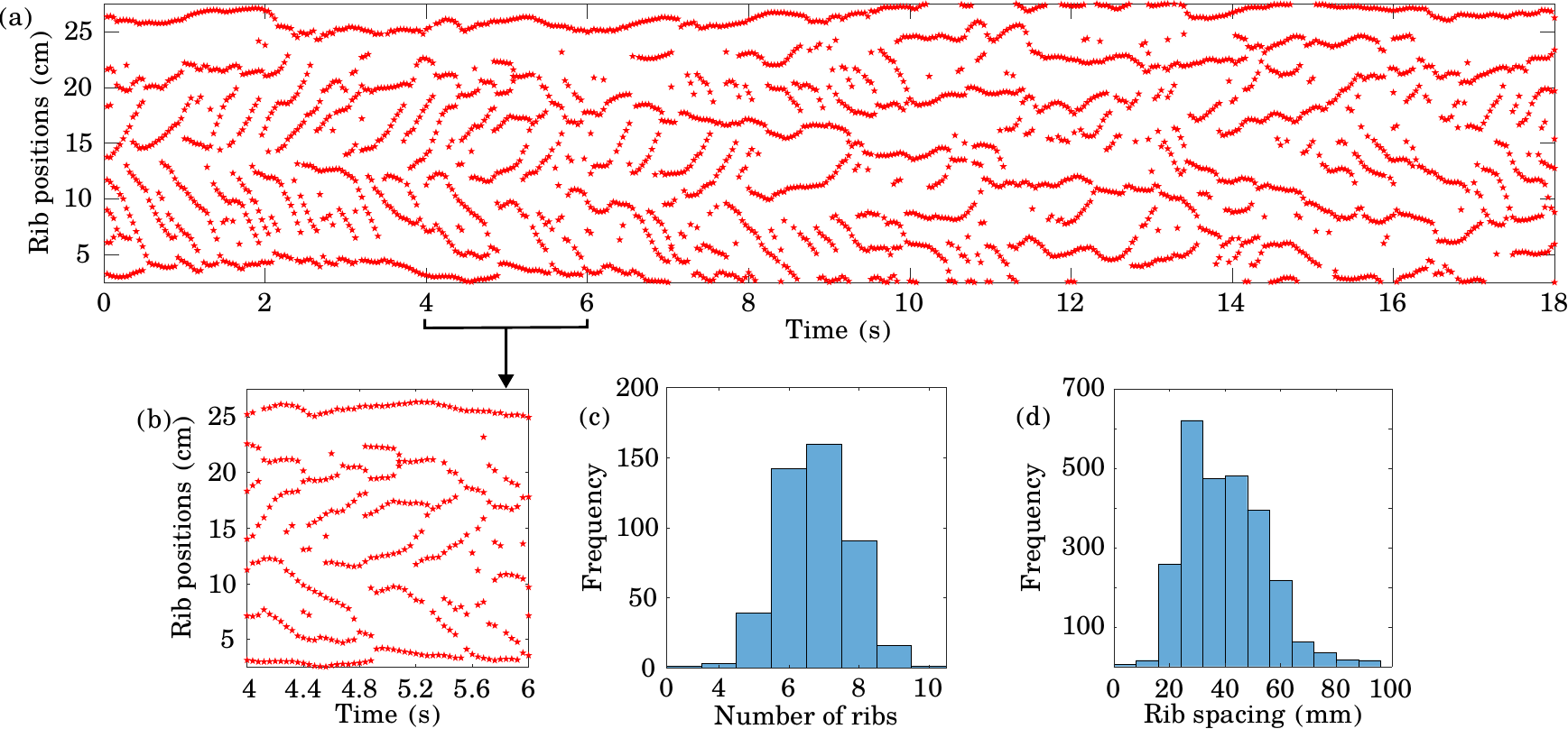}
\end{center}
\caption{(a)-(b) Spatiotemporal evolution of rib positions, $H/R=0.05$ and $\Omega = 73$~rpm, water/UCON mixture $\mu$ = $0.096$~Pa~s, frame rate $25$~images/s {(see Supplemental Video I)}.  The range of positions on the vertical axis covers the total width of the wheel, $30$~cm. (c)-(d) Histograms of number of ribs and rib spacing for the same conditions.}
\label{fig:spatio}
\end{figure}

Figure \ref{fig:spatio} shows a typical spatio-temporal diagram over $18$ seconds for the water/UCON mixture (the frame rate is 25 images/s for all acquisitions). This diagram shows that the ribs formed near the edges of the wheel are mostly pinned to the same location, while the ribs located near the center of the wheel tend to travel to the nearest edge. This process can also be observed on Fig. \ref{fig:Setup}(b), where a small rib, indicated by a white arrow, is emerging midway between both sides, as the two ribs next to it migrate to opposite sides.
\begin{figure}
\begin{center}
\includegraphics[width=1\textwidth,keepaspectratio=true]{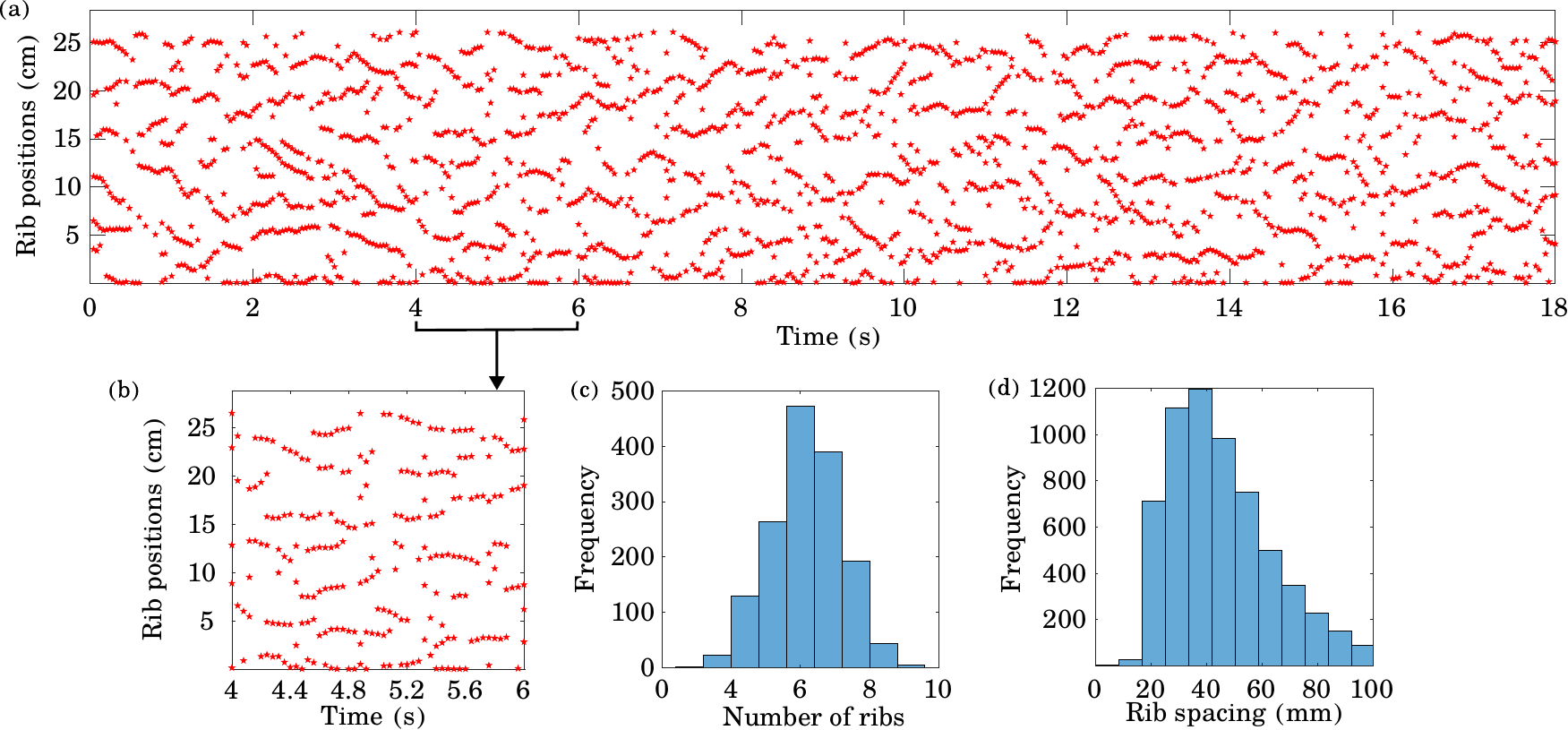}
\end{center}
\caption{(a)-(b) Spatiotemporal evolution of rib positions, $H/R=0.05$ and $\Omega = 122$~rpm, water, frame rate 25 images/s  {(see Supplemental Video I)}. (c)-(d) Histograms of number of ribs and inter-rib distances for the same conditions.}
\label{fig:spatio_w}
\end{figure}
The resulting distributions {of the total number of ribs and of distances between neighbouring ribs are} displayed in Fig. \ref{fig:spatio}(c) and \ref{fig:spatio}(d) respectively. The distribution of inter-rib distances is not symmetric. In particular, very large inter-rib distances can be observed due to outward drift of ribs as inferred from the spatio-temporal diagram.

In Fig. \ref{fig:visu_nappes}, the ribs observed with both water and the water/UCON mixture present spacings of the same order of magnitude, in spite of the very different values in viscosity for both liquids. However, the flow is strongly inertial for the case of water, for which as mentioned in the introduction the Reynolds number based on the rib thickness can be as high as $10^4$.
%{Indeed, the impact of viscosity is visible in the nature of the flow at the scale of the rib thickness.}
{Indeed, the impact of viscosity is visible on the rim of liquid sheets and the sheet smoothness in both images displayed in Fig. \ref{fig:visu_nappes}. Thus, for the less viscous case (water),  we observe that the sheet is strongly distorted, and droplets may even detach from the rib as water is ejected from the wheel {(see Supplemental Video I)}}. The corresponding spatio-temporal diagram and histograms for the number of ribs and inter-rib distances are shown in Fig. \ref{fig:spatio_w} for the case of water. The same features as for the viscous water/UCON case are again observed for the dynamics of the liquid sheets \i.e, the ribs close to the edge remain approximately pinned to their location, while ribs formed midway between the edges tend to travel to the closer side. The strong fluctuations in rib shape, which introduce noise in rib detection, render the tracking of ribs a little more difficult than in Fig. \ref{fig:spatio}(a). A satisfying convergence for typical histograms of rib spacing can nonetheless be obtained, if the recording duration is extended to $60$ seconds for the water series.

For both water and water/UCON mixtures, the histograms of the number of ribs {are} approximately symmetric, {and exhibit close to} \textit{normal} distributions. In contrast, the histograms of the inter-rib distances are asymmetric. 
%All histograms present a {single} maximum : j'ai commenté car le figure 4d a 2 maxima !!

In what follows, we chose to retain the maximum value of the latter histogram as a relevant measure of the distance between ribs for a given set of conditions, rather than the mean value which would lead to overestimate the rib spacing. Note that this value is computed from a parabolic fit based on the histogram maximum and its two adjacent values. Hereafter, this value will be referred simply as rib spacing $\lambda$, for the sake of brevity. Experiments performed by decreasing the distance between the wheel and the tank bottom down to $1$~cm did not show any impact on {either} the rib spacing {or} the above mentioned rib dynamics. Therefore, results given in the remainder of the paper were all obtained for a fixed distance of $6.3$~cm between the wheel and the tank floor, as in Fig. \ref{fig:Setup}(a).

\subsection{The most probable rib spacing}
\begin{figure}
	\begin{center}
		\includegraphics[width=1\textwidth,keepaspectratio=true]{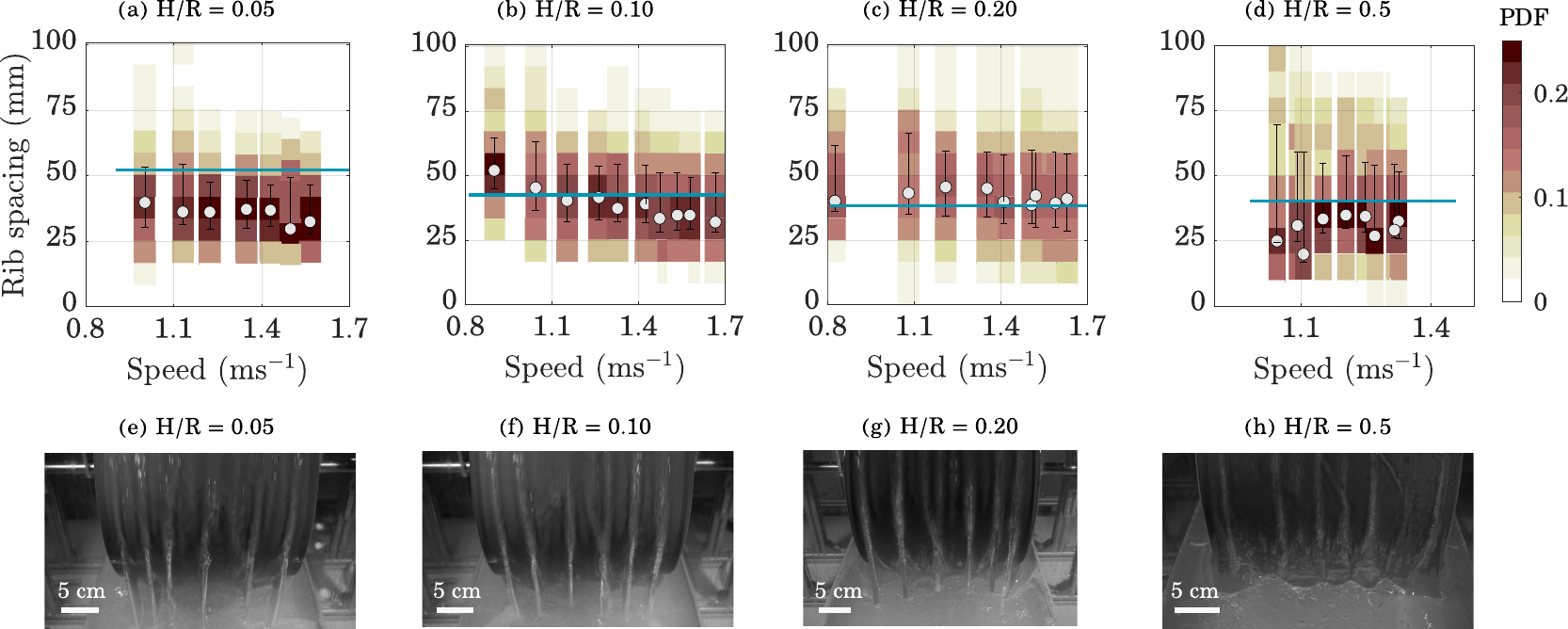}
	\end{center}
	\caption{Variations of the most probable rib spacing ($\circ$) as a function of velocity $U$, for 30\%~water/UCON mixture, and for different immersion depths $H/R$: (a) $H/R=0.05$ and $\mu = 0.096$~Pa.s, (b) $H/R=0.1$ and $\mu = 0.099$~Pa.s, (c) $H/R=0.2$ and $\mu = 0.081$~Pa.s , (d) $H/R=0.5$ and $\mu = 0.095$~Pa.s. Error bars indicate the values of the first and last quartiles. {Colourbar represents the normalised PDF values from the histogram of the inter-rib distances.} All instant photographs in (e)-(g) correspond to velocities $U = 1.4 \pm 0.1$~m/s, while it is $1.25$~m/s for (h) H/R=0.5. {Continuous line (cyan) in each figure was computed using expression (\ref{eqn:waveLen_ST}) based on \textit{directional} Saffman-Taylor instability.} See Supplemental Material for videos.}
\label{fig:rib_spacing}
\end{figure}

{We show in Fig. \ref{fig:rib_spacing} the distribution of the measured inter-rib spacing along with the maximum (symbol $\circ$) of the rib spacing histogram as a function of wheel rotation velocity, and for four immersion depths $H/R$.} The results of this figure are all for the 30\% UCON and 70\% water mixture. The small variations in viscosity between each series correspond to the variation in temperature between them. 

The measured rib spacing values are similar for all depths, of the order of $4$~cm. The lower and upper error bars indicate the values of the first and last quartile, respectively. {The main observation is that the rotation speed has no significant impact on the rib spacing.}
%When velocity is increased, a very slight decrease in the most probable rib spacing may be inferred. This trend, which is within error bars, is observed for all immersion depths, except for the larger $H/R=0.5$ case in Fig. \ref{fig:rib_spacing}(h). 
%%This different behaviour may be due to the specific conditions of that case: as already shown in \citet{jerome2021inertial}, the thickness of the ribs increases when $H/R$ is increased {(see Supplemental Videos II and III)}. 
{At $H/R=0.5$, Supplemental Videos II and III show that the rib thickness becomes of the order of magnitude of the rib spacing, which limits rib formation at low velocities, as illustrated in Fig. \ref{fig:rib_spacing}(h).} Note also that {as mentioned in the introduction} bubbles are generated when air is entrained on the plunging side of the wheel, or when the ribs fall back into the bath. They are present below the wheel and can be further entrained along with the working liquid. These bubbles tend to move into the ribs, and to regroup in lines as they are swept on the other side along the surface of the wheel. These bubbles are the reason why the upper part of the ribs appears in a lighter shade (see Fig. \ref{fig:rib_spacing}). The rib spacing values measured for $H/R=0.5$ at velocities lower than $1.25$~m/s correspond to the spacing between these bubble trails. Their values are of the same order of magnitude as for the rib spacing measured at lower $H/R$. 
\begin{figure}
	\begin{center}
		\includegraphics[width=1\textwidth,keepaspectratio=true]{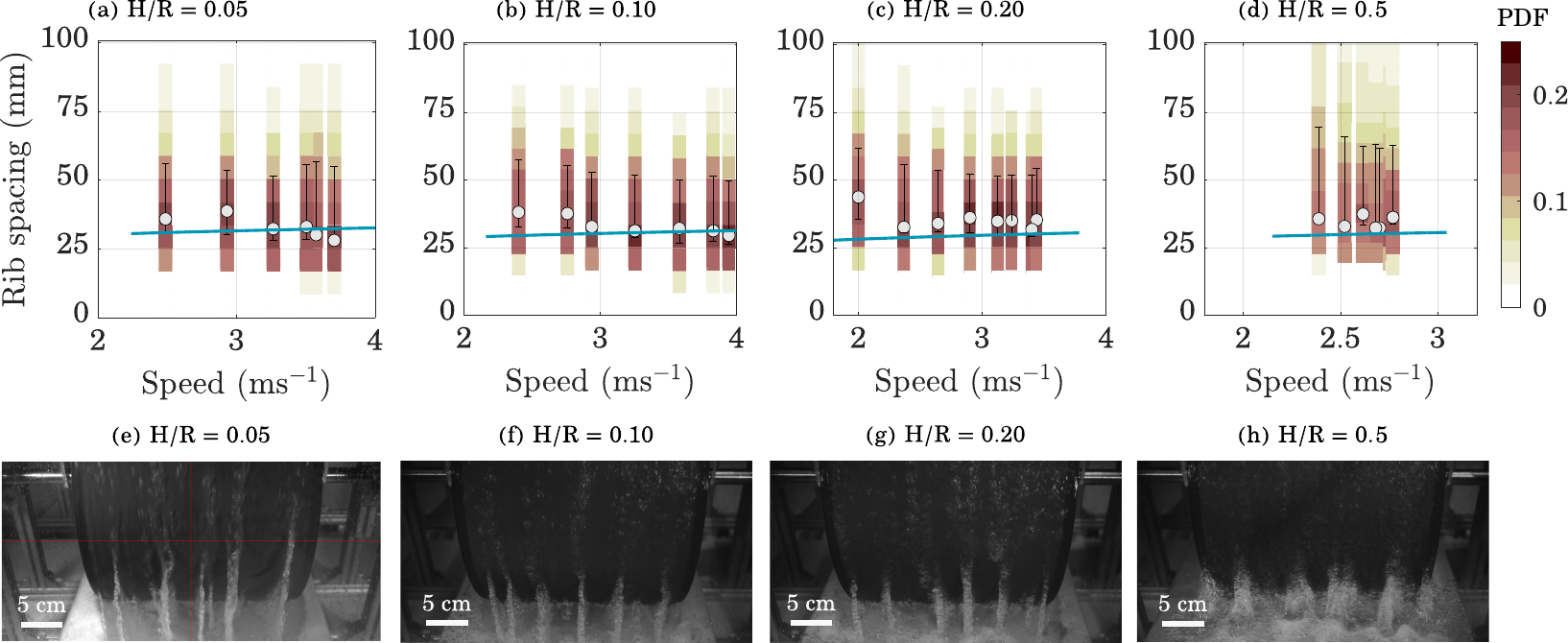}
	\end{center}
	\caption{Variations of the most probable rib spacing ($\circ$) as a function of velocity $U$, for water at different immersion depths $H/R$: (a) $H/R=0.05$, (b) $H/R=0.1$, (c) $H/R=0.2$ and (d) $H/R=0.5$. {Colourbar represents the normalised PDF values from the histogram of the inter-rib distances.} All illustration photographs (e) to (h) correspond to velocities close to 2.7~m/s. {Continuous line (cyan) in each figure was computed using expression (\ref{eqn:waveLen_ST}) based on \textit{directional} Saffman-Taylor instability.}}
\label{fig:rib_spacing_w}
\end{figure}
Fig. \ref{fig:rib_spacing_w} presents the evolution of the most probable rib spacing as a function of velocity for experiments carried out with water. {For the sake of completeness, the inter-rib spacing distribution is also provided for each case investigated here.} The measured values are close to, but slightly smaller than the values for the water/UCON mixture, with typical rib spacings of the order of $3.5$~cm. The same slight decrease of rib spacing with velocity is observed for the three smaller immersion depths. This decrease is observed in the position of the most probable spacing (symbol $\circ$), but also in the position of the first and fourth quartile, which are slightly shifted to lower values when speed is increased. As observed for the water/UCON mixture, no significant influence of the immersion depth is observed on the rib spacing in Fig. \ref{fig:rib_spacing_w}(e) - (h), even though the thickness of the ribs varies dramatically between $H/R=0.05$ and $H/R=0.5$. {Note that the effect of surface roughness, using for example a bubble wrap (not shown here), hardly changed these observations.}

\section{Mechanism of Pattern Formation}
\label{sec:STmechanism}
{As discussed in the introduction, the observed ribbing patterns are very similar to those already observed for the Rayleigh-Taylor (RT) instability of a thin liquid films falling under gravity, or under centrifugal acceleration. They also resemble those observed in directional Saffman-Taylor (ST) instability of wall-entrained viscous liquid in thin diverging gaps. The maximum growth rate of infinitesimal perturbations in the RT case is well known \citep{Vrij1966, Babchin1983, Fermigier1992, Mikaelian1996, Balestra2018}:
\begin{eqnarray}
%%	\omega_i^{RT} \simeq \dfrac{\rho^2 g_a^2 \delta_f^3}{4\mu \sigma},
	\omega_i^{RT} \sim \dfrac{\rho^2 g_a^2 \delta_f^3}{\mu \sigma},
	\label{eqn:omg_RT}
\end{eqnarray}
where $\delta_f$ is the film thickness, and $g_a$ is the apparent acceleration equal to the effective acceleration due to gravity, or $U^2/R$. In the ST case, its expression from \citep{Hakim1990, rabaud1994dynamiques} can be simplified as follow as
\begin{eqnarray}
	\omega_i^{ST} \sim \left(\dfrac{U}{\delta_f}\right) Ca^{1/2},
%%	\omega_i^{ST} \simeq \left(\dfrac{U}{\delta_f}\right) Ca^{1/2},
	\label{eqn:omg_ST}
\end{eqnarray}
where the driving pressure gradient of the base flow has been taken as $\mu U/\delta_f^2$. For water, with $g = 9.81$~m~s$^{-2}$ and $\delta_f \sim \sqrt{\mu U/\rho g}Ca^{1/6}$ \citep{levich1942dragging, wilson1982drag}, we get $\omega_i^{ST} \sim 750 $~s$^{-1}$ while $\omega_i^{RT} \sim 5$~s$^{-1}$. For the viscous UCON/water mixtures in our experiments $Ca > 1$ and so, we take $\delta_f \sim 2/3 \sqrt{\mu U/\rho g}$ \citep{derjagin1943thickness, JinAcrivosMunchPoF2005}. Then, the order of magnitude of growth rates are $550$~s$^{-1}$ and $130$~s$^{-1}$, respectively, in the case of linear ST and RT instability.}

{These values suggest that the \textit{directional} ST mechanism is perhaps the most suited for the observed patterns in our experiments. Nonetheless, using $g_a = U^2/R$ when $U = 2$~m~s$^{-1}$ speed quadruples the RT growth rate. Also, the RT growth rate is not very small when compared with that of the ST case for the UCON/water mixtures even at speeds $U \approx 1$~m~s$^{-1}$ and $g_a = 9.81$~m~s$^{-2}$. So, the RT mechanism might not be negligible at high-speeds, for the viscous case in our study. However, the rib spacing did not evolve considerably when we doubled (UCON/water), or even quadrupled (water), the drum speed. Furthermore, advection is strong in our experiments, contrary to the case of a falling liquid film.}

Thus, following previous investigations on ribbing patterns in thin film-splitting coating flows and rimming flows \citep{pitts1961flow, mill1967formation, adachi1988coating, coyle1990stability, Hakim1990, rabaud1994dynamiques}, we expect that the driving mechanism in our study is due to an adverse pressure gradient on the rising side of the rotating drum. We now proceed to demonstrate the existence of this adverse pressure gradient in the rotary drag-out flow, and then relate its magnitude to the characteristic wavelength of a linear \textit{directional} ST instability. For the sake of simplicity, the analysis is restricted to the 2$D$ inclined plate drag-out problem.

\subsection{Pressure gradient in the drag-out problem}
\label{subsec:advP}
\begin{figure}
	\begin{center}
		\includegraphics[width=1\textwidth,keepaspectratio=true]{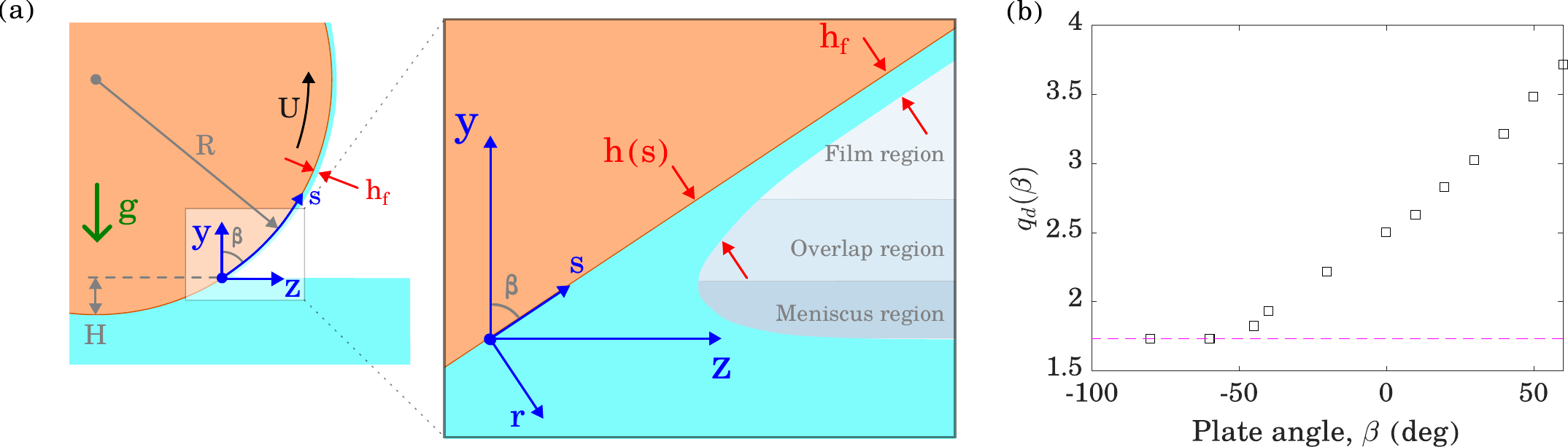}
	\end{center}
	\caption{{(a) Schematic of the rotary drag-out problem and its equivalent inclined flat plate LLD flow where {$\beta = \sin^{-1}(1-H/R)$} with $s$ the curvilinear coordinate such that $s = 0$ at the point of intersection between the liquid level at rest and the wheel. Note that $h(s)$ in the overlap region generally decreases as $s > 0$ increases down to the asymptotic value $h_f$ in the fully-developed film region. (b) The second prefactor $q_d(\beta)$ in the expression (\ref{eqn:hf_lubri_A1}) is estimated from the maximum flow rate condition via the Stokes equation at $Ca >> 1$  \cite[Fig. 8]{JinAcrivosMunchPoF2005}.}}
	\label{fig:rotaryLLD_A1}
\end{figure}

Consider the $2$D rotary drag-out problem as illustrated in Fig. \ref{fig:rotaryLLD_A1}(a) wherein an infinitely-long {horizontal} rotating cylinder entrains a liquid out of a bath. Similar to the classical  {Landau-Levich-Deryaguin (LLD)} \citep{levich1942dragging, derjagin1943thickness} flat-plate drag-out, the flow field can be distinguished into a fully-developed film region and a meniscus region separated by an overlap region. Also, when the local film thickness $h(s) \ll R$, {the velocity component, say $u$, along the curvilinear coordinate $s$} is dominated by viscous shear. So, in the film and the overlap region, $2$D lubrication theory should hold as long as the flow is quasi parallel and the local Reynolds number based on the film thickness $Re_f = \left(\rho g R/\mu\right) \left(h_f/R\right)^2$ is small. Furthermore, as suggested by \citet{wilson1982drag}, if the dynamic meniscus length is much  smaller than the cylinder radius, it is possible to reduce the rotary drag-out to an equivalent inclined flat plate entrainment flow such that the plate angle with respect to the vertical is given by {$\sin \beta = 1 - H/R$}, where $H$ is the immersion depth as depicted in Fig. \ref{fig:rotaryLLD_A1}(a). %{Under these assumptions, local liquid speed parallel to the plate should be}
{Under the lubrication approximation
\begin{eqnarray}
	0 &= & -\rho g \cos \beta - \dfrac{\partial p}{\partial s} + \mu \dfrac{\partial^2 u}{\partial r^2} + \mathcal{O} \left(Re_f\right), \\
	0 &= & \rho g \sin \beta - \dfrac{\partial p}{\partial r} + \mathcal{O} \left(Re_f\right),%\\
%%	0 &= & - \dfrac{\partial p}{\partial x} + \mu \dfrac{\partial^2 w}{\partial x^2} + \mathcal{O} \left(Re_h\right),
	\label{eqn:NS_lubri}
\end{eqnarray}
where $p$ is the liquid pressure in the film. {The boundary conditions are $u = U$ at the wall ($r = 0$), and $\partial u/ \partial r = 0$ at $r = h(s)$.} In the absence of any axial pressure gradient, the inclined LLD flow at first-order is}

\begin{eqnarray}
	u(r, s) = \dfrac{1}{2\mu} \left( \dfrac{\partial p}{\partial s} + \rho \tilde{g} \right) \left(r^2 - 2h r \right) + U,
	\label{eqn:u_lubri}
\end{eqnarray}
where $\tilde{g} = g \cos \beta$ is the effective gravity. Now, by applying mass conservation \.i.e,
\begin{eqnarray}
	\dfrac{d}{ds} \left(\int_{r=0}^{r = h(s)} u(r, s) dr\right) & = 0,
	%\Rightarrow G \equiv \dfrac{\partial p}{\partial s} &= & \rho \tilde{g} \left[ \left( a^2 - \left(a^2 - 1 \right)\dfrac{h_f}{h}\right) \dfrac{h_f^2}{h^2} - 1 \right],
	\label{eqn:mass_conv_lubri_A1}
\end{eqnarray}
between the overlap region and the film region (where $h(s) = h_f$ and $p(s, r) = P_0$, the atmospheric pressure) we get
\begin{eqnarray}
	-\dfrac{h^3}{3\mu} \left( \dfrac{\partial p}{\partial s} + \rho \tilde{g} \right) + Uh &= &-\dfrac{h_f^3}{3\mu} \rho \tilde{g} + Uh_f,\notag \\
	\Rightarrow \dfrac{\partial p}{\partial s} &= &-\rho \tilde{g} \left(1 - \dfrac{h_f^3}{h^3} \right) + \dfrac{3 \mu U}{h^2} \left(1 - \dfrac{h_f}{h} \right),
	\label{eqn:gradP_frm_masse_conv_A1}
\end{eqnarray}
{which can be rewritten after some algebra, and by defining the pressure gradient $G = \partial p/\partial s$, as follows
\begin{eqnarray}
	G &= &- \rho \tilde{g} \left(1 - \dfrac{h_f}{h}\right)\left( 1 + \dfrac{h_f}{h} -\left(a^2 - 1 \right) \dfrac{h_f^2}{h^2} \right),
	\label{eqn:gradP_lubri_A1}
\end{eqnarray} 
where parameter $a = \sqrt{3} Ca^{1/2}l_c/h_f$ with $l_c = \sqrt{\sigma/\rho \tilde{g}}$ the characteristic capillary length scale. {It is customary to further relate the pressure gradient to the local film curvature via the Young-Laplace theorem and then obtain an expression for the film thickness $h(s)$. Thereupon, it is possible to obtain both the flow and pressure field \citep{WeinsteinChemEngg2001}. It is also possible to include inertial effects \citep{Cerro1980rapid, Kheshgi1992rising, WeinsteinRuschak_AnnRevFluMech2004}. In our formulation, however, all the details of the flow field are resumed in parameter $a$, which contains the film-thickness $h_f$.  We estimate this parameter from the classical formulae for $h_f$ under lubrication approximations.} At small Capillary numbers ($Ca \ll 1$), the film thickness is given by
\begin{eqnarray}
h_f = \dfrac{0.9458 ~ l_c Ca^{2/3}}{\sqrt{\sec \beta + \tan \beta}} \left(1 - \dfrac{0.113}{{\sec \beta + \tan \beta}}  Ca^{1/3} + \mathcal{O}(Ca^{2/3}) \right),
	\label{eqn:hf_lubri_A1}
\end{eqnarray}
which is the generalization of Landau-Levich vertical drag-out flow \citep{levich1942dragging} to the case of an inclined flat plate for different $0 \leq \beta < \pi/2$ by  \cite{wilson1982drag}. At $Ca \gg 1$, $h_f \propto l_c Ca^{1/2}$ where the constant of proportionality depends on the plate inclination angle ($\beta$), as established by \citet[see Fig. 8]{JinAcrivosMunchPoF2005}\footnote{This was first hypothesized by \citet{deryagin1964film} for the case of a vertical drag-out flow.} using Direct Numerical Simulations of the Stokes equation for the inclined-plate drag-out problem. Therefore, we obtain
\begin{eqnarray}
a = \left\{
        \begin{array}{ll}
            %\dfrac{0.9458 ~ l_c Ca^{2/3}}{\sqrt{1 + \sin \beta}} \left(1 - \dfrac{0.113 \cos \beta}{{1 + \sin \beta}}  Ca^{1/3} + \mathcal{O}(Ca^{2/3}) \right) & \forall Ca \ll 1, \\
            {q_w(\beta) ~ Ca^{-1/6}} + \mathcal{O}\left( Ca^{1/6} \right) & \forall Ca \ll 1, \\
            	& \\
            q_d(\beta) & \forall Ca \gtrsim 1,
        \end{array}
    \right.
	\label{eqn:a_lubri_A1}
\end{eqnarray}
such that $q_w(\beta) = 1.06 \sqrt{3\left(\sec \beta + \tan \beta \right)}$ and $q_d(\beta) \geq \sqrt{3}$ as presented in Fig. \ref{fig:rotaryLLD_A1}(b).}

\begin{figure}
	\begin{center}
		\includegraphics[width=1\textwidth,keepaspectratio=true]{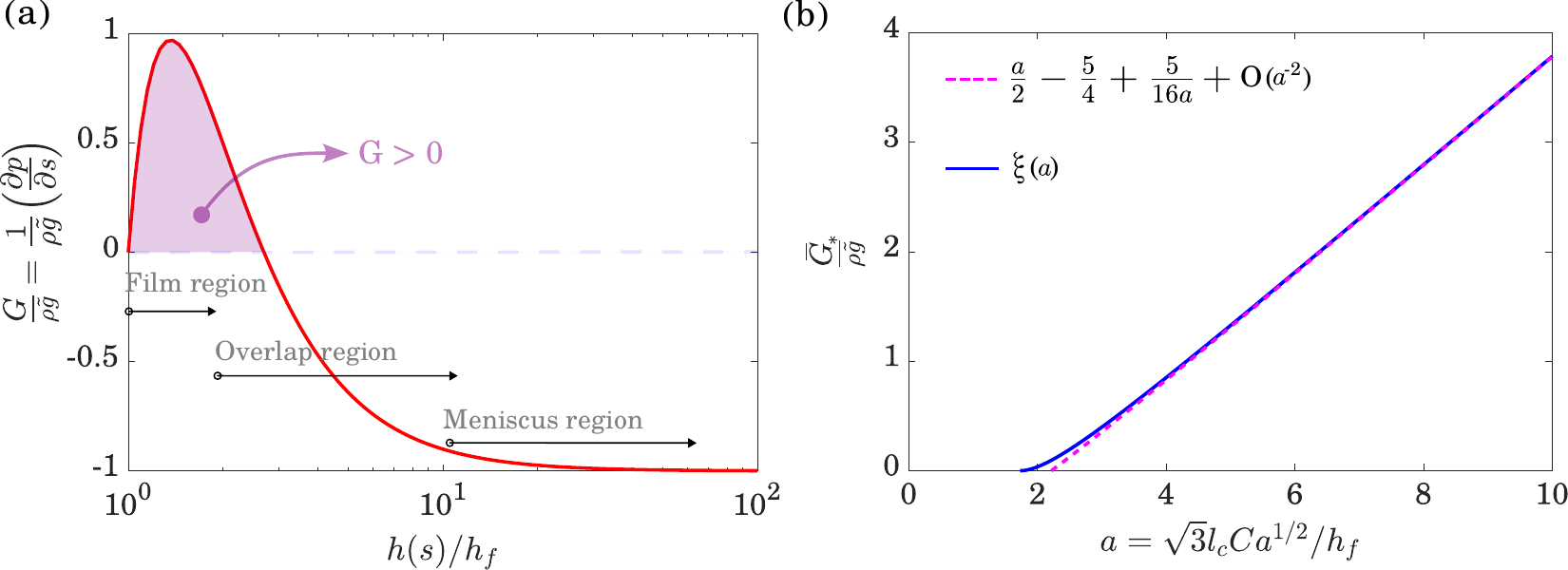}
	\end{center}
	\caption{{(a) Non-dimensional pressure gradient along the inclined flat plate as a function of the relative local film thickness $h(s)/h_f$ as computed from (\ref{eqn:gradP_lubri_A1}) with $a = 4$. (b) The average non-dimensional adverse gradient pressure is presented against the parameter $a > \sqrt{3}$ shows that the series expansion of $a$ in (\ref{eqn:gradPavg_lubri_A1}) compares well with the computed value from (\ref{eqn:gradPavg_defn_A1}).}}
	\label{fig:rotaryLLD}
\end{figure}
{In Fig. \ref{fig:rotaryLLD}(a), we present a typical evolution of the pressure gradient as a function of the normalised film thickness for $a = 4$. Firstly, in the meniscus region the relative film thickness $\tilde{h}(s) = h(s)/h_f \ll 1$, the pressure gradient here tends to its hydrostatic limit {$G=-\rho \tilde{g}$}. As the curvilinear coordinate $s > 0$ increases, in the overlap zone, the pressure gradient increases as well, since $h(s)$ decreases monotonically. Figure \ref{fig:rotaryLLD}(a) shows that $G$ reaches a maximum before decreasing to zero in the film region, where the film thickness $h(s) = h_f$. This implies that there exists a region of positive pressure gradient for the condition $a = 4$. More generally, in the overlap region adjacent to the film region shown in Fig. \ref{fig:rotaryLLD_A1}(a), we can admit that $h(s) \sim h_f \left( 1 + \epsilon \right)$, for some $\epsilon \ll 1$. Then, we find that 
\begin{eqnarray}
	G \sim \rho \tilde{g} (a^2 - 3) \epsilon + \mathcal{O}(\epsilon^2).
	\label{fig:gradP_overlapZone_A1}
\end{eqnarray}
This pressure gradient is of $\mathcal{O}(\epsilon)$ and positive if $a > \sqrt{3}$ (or in other words, $h_f < l_c Ca^{1/2}$). Thus, a necessary condition for an adverse pressure gradient to exist in the overlap zone is that the film-thickness is larger than the Jeffreys drainage lengthscale $l_c Ca^{1/2} \equiv \sqrt{\mu U/\rho \tilde{g}}$ \citep{jeffreys_1930}. Furthermore, expression (\ref{eqn:gradP_lubri_A1}) indicates that $G(s) > 0$ as long as $h_f < h(s) < h_{\ast}$, where ${h_{\ast}} = \left( \sqrt{4 a^2 - 3} - 1\right)h_f/2$ is the film-thickness at some location upstream of the film region ($h(s) \gtrsim h_f$). Thus, the pressure gradient $G$ is positive for all $h_f < h(s) < h_{\ast}$. Thereby, we propose to estimate the average adverse pressure gradient in this zone as follows 
\begin{equation}
\bar{G}_{\ast} = \dfrac{1}{\left(h_{\ast} - h_f\right)} \int_{h_f}^{h_{\ast}} G(h) dh \equiv \rho \tilde{g} \xi(a),
	\label{eqn:gradPavg_defn_A1}
\end{equation}
where $\xi(a)$ is some irrational function that increases monotonically with the parameter $a > \sqrt{3}$ as presented in Fig. \ref{fig:rotaryLLD}(b). The general expression of $\xi(a)$ can be developed as a Laurent Series
\begin{equation}
	\xi(a) = \dfrac{a}{2} - \dfrac{5}{4} + \dfrac{5}{16a} + \mathcal{O}\left(\dfrac{1}{a^2}\right) > 0,
	\label{eqn:gradPavg_lubri_A1}
\end{equation}
at sufficiently large values of $a$ which, as depicted in Fig. \ref{fig:rotaryLLD}(b), is a good approximation for $\xi(a)$ when $a \gtrsim 3$. 

In summary, an adverse pressure gradient such as the one evidenced in Fig. \ref{fig:rotaryLLD}(a) will exist} in the neighbourhood close to the film region ($h(s) \gtrsim h_f$) provided $a > \sqrt{3}$. Its average scales as the absolute value of the hydrostatic pressure gradient $\rho \tilde{g}$ and some function of plate angle $\beta$ and the Capillary number $Ca$. {The existence of an adverse pressure gradient could also be qualitatively inferred from the shape of the meniscus between the reservoir and the film region in a drag-out flow. However, the previous analysis does not rely upon any estimate of the meniscus curvature. Here, the pressure gradient arises directly out of a dynamic equilibrium in the presence of viscous shear, gravity and capillarity. The effect of capillarity is contained in the chosen model for film thickness $h_f$.}

\subsection{Directional Saffman-Taylor instability in rotary drag-out flows}
{The existence of a region of positive pressure gradient between the film and the overlap regions in the lubrication limit leads to the so-called \textit{directional} Saffman-Taylor instability \citep{Hakim1990}. Note that the Saffman-Taylor (ST) instability in a porous medium, or a Hele-Shaw cell, refers to the situation when a fluid of viscosity $\mu_1$ is displaced by another less viscous fluid when $\mu_2 < \mu_1$ \citep{SaffmanTaylor1958}. Whereas, in the classical ST instability, the interface between the two immiscible fluids is under displacement towards the more viscous fluid, the meniscus in the LLD drag-out flow is immobile in the laboratory frame. However, in the frame of reference of the moving plate, the air/liquid interface can be seen as an advancing meniscus towards the more viscous fluid \i.e., water or water/UCON mixtures in LLD flow. Furthermore, the pressure gradient is favourable as one moves across the interface from the less viscous fluid to the more viscous fluid. This is also the case for the drag-out problem in the reference frame attached to the moving plate. It is precisely in this sense that we propose the \textit{directional} Saffman-Taylor mechanism in the overlap region for the development of ribbing in inertial rotary drag-out flows.}

{For the case of a liquid of viscosity $\mu$ displaced by air, the most unstable wavelength $\lambda_{ST}$ of the Saffman-Taylor instability is given by $\lambda_{ST} \simeq \pi b \sqrt{\sigma/ \mu V}$, where the viscosity of air has been neglected in comparison with that of the liquid, $b$ is the thin gap between two horizontal plates and $V$ is the speed at which the air/liquid interface advances \citep{SaffmanTaylor1958, Chuoke1959, Homsy1987}. This expression can be rewritten in terms of the driving pressure gradient, say $G$, as $\lambda_{ST} \simeq 2\pi \sqrt{3\sigma/\vert G \vert}$, since $G =  -\mu V/ 12 b^2$. In our case for the LLD flow, we expect that the wavelength $\lambda_{\ast}$ of the most unstable axial perturbation is given by the average pressure gradient {$\bar{G}_{\ast}$}  in the neighbourhood of the film region where the pressure gradient is positive. So, we obtain
\begin{equation}
	\lambda_{\ast} = 2 \pi l_c\sqrt{3/\xi(a)},
	\label{eqn:waveLen_ST}
\end{equation}
where $\xi(a)$ is the normalised average adverse pressure gradient in the overlap zone of the rotary drag-out problem as defined in (\ref{eqn:gradPavg_defn_A1}), and $l_c = \sqrt{\sigma/\rho \tilde{g}}$ is the characteristic capillary length scale. By using the series expansion of $\xi(a)$ \.i.e, (\ref{eqn:gradPavg_lubri_A1}) along with the lubrication approximation (\ref{eqn:a_lubri_A1}) for the parameter $a = \sqrt{3} l_c Ca^{1/2}/h_f$  in the above relation, we estimate that $\lambda_{\ast} \sim 2 \pi l_c \sqrt{3/q_w(\beta)}  Ca^{1/12}$ at low Capillary numbers $Ca \ll 1$ and $\lambda_{\ast} \sim 2 \pi l_c \sqrt{3/q_d(\beta)}$ at $Ca \gg 1$, up to $\mathcal{O}\left(a^{-2}\right)$. {This implies that the most-unstable wavelength is expected to vary only very weakly with the Capillary number $Ca = \mu U/ \sigma$, and hence with velocity. This is consistent with the experimental observations that velocity has almost no impact on the rib spacing.} This also means that the liquid density $\rho$ and surface tension $\sigma$ via the capillary length $l_c$ must play an important role in the rib spacing.}

{Finally, we point out that the above expression (\ref{eqn:waveLen_ST}) does not account for the spatial variation of the pressure gradient along the flow direction \.i.e., the curvilinear coordinate $s$. Also, in such a case where the pressure gradient is non-uniform, \textit{linear} instability characteristics should be a function of the flow geometry. In fact, this has been done for the case of the \textit{Printer's instability} \citep{coyle1990stability, Hakim1990, rabaud1994dynamiques, Reinelt_DirectionalViscFingering_JFM1995}. So, in order to build a more rigorous estimate of the wavelength for the case of the rotary LLD flow, a \textit{global} stability analysis \citep{Chomaz2005, Theofilis2011} is necessary. Even though this deserves further investigations, it is beyond the scope of the present work. Furthermore, the analysis should be adapted to incorporate liquid inertia since $Re_f$ is rarely small in our experiments where ribbing is observed. Nevertheless, in the following, we propose to study the relevance of the estimation provided in (\ref{eqn:waveLen_ST}) for the most-unstable wavelength by comparing it with experiments and simulations.}

{Continuous lines (cyan) in Fig. \ref{fig:rib_spacing} and Fig. \ref{fig:rib_spacing_w} are obtained using the above formula where the average adverse pressure gradient $\bar{G}_{\ast}$ is computed via (\ref{eqn:a_lubri_A1}) \& (\ref{eqn:gradPavg_defn_A1}) based on the lubrication approximation \citep{levich1942dragging, derjagin1943thickness, wilson1982drag, JinAcrivosMunchPoF2005}. In our experiments, the capillary number is such that $0.03 < Ca < 0.06$ for water and $1.4 < Ca < 2.8$ for the water/UCON mixture. Within these ranges, the non-dimensional parameter $a = \sqrt{3} l_c Ca^{1/2}/h_f$ in expression (\ref{eqn:a_lubri_A1}) varies only between $2$ and $4$. For all data points in water/UCON mixtures (Fig. \ref{fig:rib_spacing}) and also, in water (Fig. \ref{fig:rib_spacing_w}), the proposed wavelength $\lambda_{\ast}$ provides a good order of magnitude despite the variations in drum immersion $H/R$ and drum speed $U$. } This is all the more interesting from the fact that a basic Saffman-Taylor analogy which includes only the average pressure gradient $G_\ast$ from calculations based on an idealized $2$D lubrication model is quite relevant to predict the most-probable rib spacing $\lambda$ in our experiments wherein the flow is strongly inertial, unsteady, spatially non-uniform and three-dimensional. %{(et inertiel ?)}

%{est-ce que ça vaudrait pas le coup de donner un odg de $\xi(a)$ afin de proposer une version plus commerciale  de 3.5? }

\section{Numerical experiments}
\label{sec:numMethods}
\begin{figure}
\begin{center}
\includegraphics[width=1\textwidth,keepaspectratio=true]{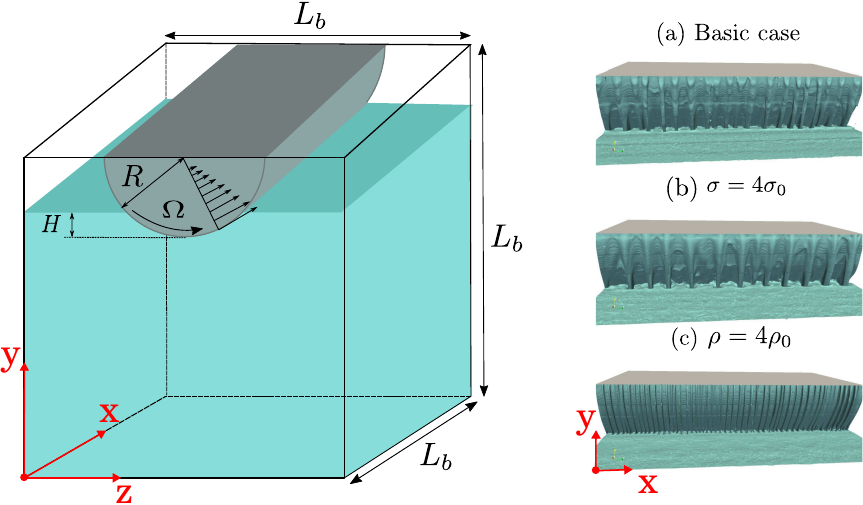}
\end{center}
\caption{Computational domain and instantaneous ribbing patterns for three case studies using \textit{open-source} DNS, namely, Basilisk. The box size is $L_b = 80$~cm and the cylinder radius is $20$~cm. See Supplemental Video IV for time evolution of the VOF for the three cases.}
\label{fig:configBasilisk}
\end{figure}

The validity of the proposed scaling for the most-probable rib spacing $\lambda$ can be further elucidated via Direct Numerical Simulations (DNS) of the rotating wheel configuration. To do so, the {transient} three-dimensional, two-phase interfacial, incompressible Navier-Stokes equations are solved with the \textit{open-source} software Basilisk \citep{popinet2009accurate,basiliskWebsite}. A spatial adaptive octree grid, particularly well suited to the multi-scale nature of the problem, is used. {The maximum level of mesh refinement, noted here as $N$, refers to an equivalent Cartesian grid resolution having $2^N$ grid points in each direction. Here $N$ varies between 9 and 10, which corresponds to equivalent $512^3$ to $1024^3$ points on a fixed grid.} The interface between the two immiscible fluids is tracked with a sharp geometric Volume-Of-Fluid (VOF) method. Basilisk is well designed for DNS of open inertial two-phase flows with large interface deformations \citep{zhang2020modeling, mostert2022high} where realistic density and viscosity ratios, such as between water and air, can be considered. The computational domain is depicted in Fig. \ref{fig:configBasilisk} (left). It consists of a cubic box of size $L_b = 80$~cm and a half cylinder of radius $R = 20$~cm and length $L_b = 80$~cm. The default symmetry boundary conditions are applied on all faces of the domain except the top face where a free outflow boundary condition is imposed. Wheel rotation is accounted for in the calculation by a penalty method in which a solid-body rotation of angular velocity $\Omega = U/R= 65$~rpm is forced on the half-cylinder. {Note that the system is left to evolve freely without any initial perturbation but under the influence of numerical noise which is of the order of the smallest spatial step size. Initial perturbations of the wetting line as a controlled disturbance were tested as well (not shown here) for wavelengths, up to the order of $L_b$ (the box size). The wavelengths obtained for the sheets were identical to those obtained without initial disturbance of the wetting line. {See Supplemental Video IV for the time evolution of interface from simulation data. The video shows that the ribs appear quickly as the liquid starts to rise along the wheel: then as the film slows down and thickens, a ‘top-down’ back flow starts to develop, which eventually plunges into the pool at the location of the ribs.}} {Note that we were unable to realise simulations for longer than 1 second in the best of the cases due to increase in computation time resulting from many drops and bubbles in the box. {At the end of the simulation (see Movie IV), once the ribs had developed, we discern a back flow along the ribs. Coarse-grained full cylinder simulations (not shown here) confirmed that this is not related to the liquid outflow boundary condition on the half-cylinder. As already shown for the case of a single rib for the case of a horizontal rotating disc \citep{jerome2021inertial}, this should in turn influence the entrained flow over the cylinder downstream of the ribs.} {Note that the finite size of the cylinder length in our experimental setup promotes rib undulations and meandering. This is further accentuated by the presence of droplets and bubbles in experiments. These effects are less important in our simulations, where symmetrical boundary condition is imposed at cylinder ends.} We recall that DNS was used to compute characteristics rib spacing and growth rates for various liquid properties. The non-linear effects due to end and top boundary conditions are not the scope of the present numerical study.}

A reference case, termed as basic case hereafter, was first run to compare the observed most-probable inter-rib distance $\lambda$ for the case of the water/UCON mixture at $U = 1.36$~m~s$^{-1}$ and $H/R = 0.2$. Here, the liquid density, viscosity and surface tension are $\rho_{0} = 1044$~kg~m$^{-3}$, $\mu_{0} = 100$~Pa~s and $\sigma_{0} = 0.05$~N~m$^{-1}$, respectively. {Note that these physical parameters correspond precisely to one of the experimental runs, see Fig. \ref{fig:rib_spacing}(g).}
%{(Dire que le basic case numérique est aussi traité expérimentalement et pourra servir de comparaison ?)} %A spatial FFT of the VOF interface in the axial direction can now be compared with the observed maximum rib spacing and the model. This is provided in Fig. \ref{fig:WaveLengthComp} with the symbols $\triangleleft$ and $\triangledown$ which correspond to two different grid resolutions. Note that the colormap indicates the simulation time in seconds. Once again a large scatter is observed about $\lambda_\ast = \lambda_0$. Nonetheless, simulations show a reasonably good match not only with  experiments but also with the Saffman-Taylor wavelength prediction. Note that the agreement is better at larger simulation times. But computational cost and resolution of a fully two phase flow due to bubbles and drops render much longer simulations quite a difficult task.
Furthermore, two sets of numerical experiments were treated in order to test the proposed scaling with the capillary length $l_c$. For these cases, we modified either the liquid surface tension to $\sigma = 4\sigma_0$, or the liquid density to $\rho = 4\rho_0$, while keeping all other parameters identical to the basic case.
{The computed iso-surfaces for VOF~$= 0.5$ are shown in Fig. \ref{fig:configBasilisk}(a)--(c).} All cases present a large number of ribs along the cylinder axis while the ribbing patterns vary between the different cases studied here. About 52 ribs are observed in the case where the liquid density is exactly $4$ times the basic case \i.e., $\rho = 4\rho_0$. This is just approximately twice the number of ribs in the basic case, and about four times that obtained in the case when only the liquid surface tension is modified to $\sigma = 4\sigma_0$ {(see Supplemental Videos IV)}. These observations show a good qualitative agreement with the proposed scaling for the rib spacing, since the latter scales with the capillary length $l_c = \sqrt{\sigma/\rho \tilde{g}}$ as indicated by  expression (\ref{eqn:waveLen_ST}). %Fig. \ref{fig:WaveLengthComp} also provides numerical results for the cases when these properties are quadrupled with respect to the above mentioned reference case. The ribbing patterns presented in Fig. \ref{fig:configBasilisk}(b) and (c) illustrate that the number of ribs is approximately doubled when the corresponding liquid density $\rho = 4\rho_{Ucon}$ and is respectively, halved when the liquid surface tension $\sigma = 4\sigma_{Ucon}$. The maximum rib spacing is reported in Fig. \ref{fig:WaveLengthComp} and they match reasonably with the expression \ref{eqn:waveLen_ST}. 

\begin{figure}
\begin{center}
\includegraphics[width=1\textwidth,keepaspectratio=true]{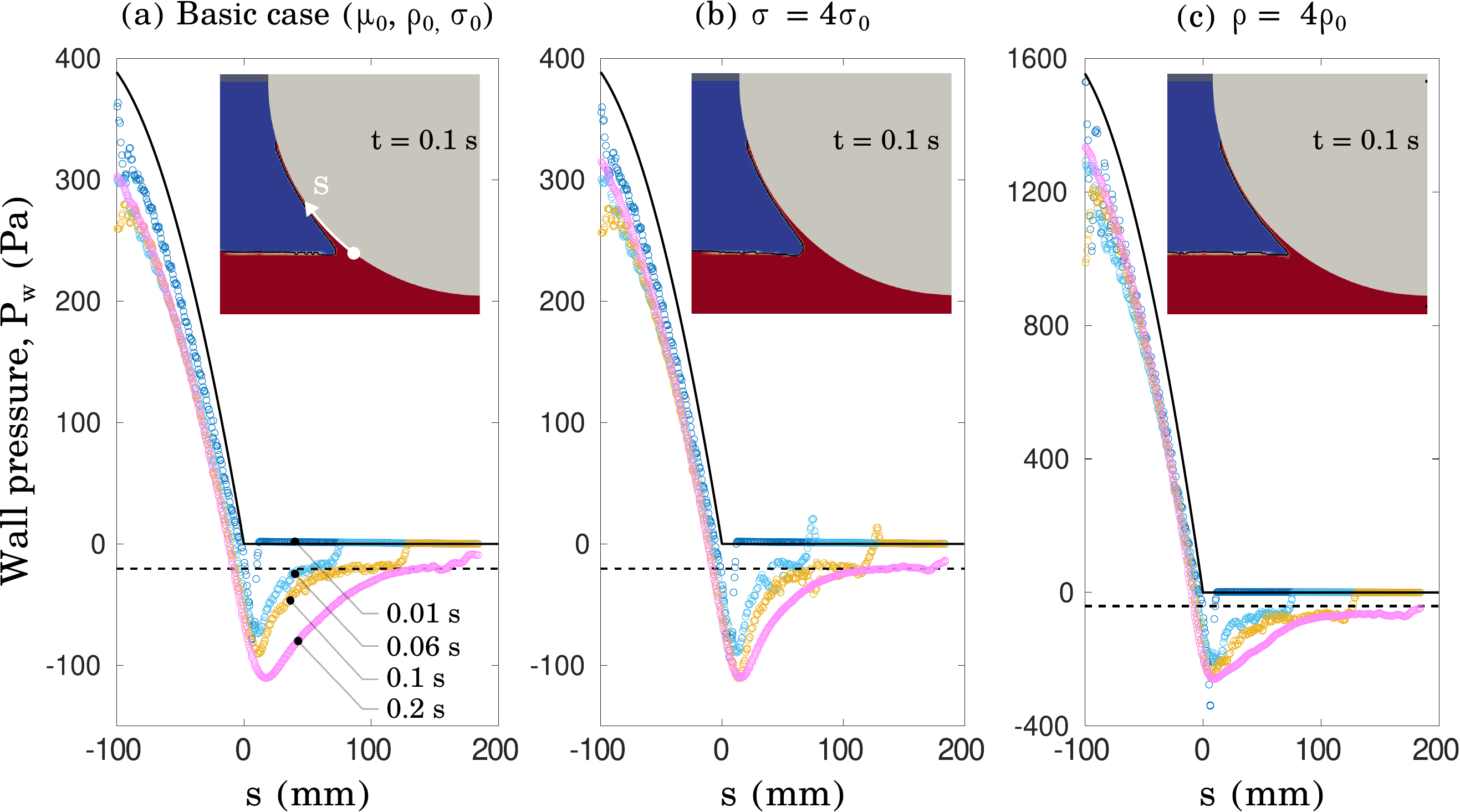}
\end{center}
\caption{Pressure profiles along the curvilinear coordinate, s within the entrained liquid film in the midplane ($x = 40$~cm), at four different timesteps: (a) Basic case ; (b) $\sigma = 4\sigma_0$ ; (c) $\rho = 4\rho_0$. {The continuous and dashed lines (black) correspond respectively to the hydrostatic case when the wheel is at rest and to the radial pressure gradient $\rho U^2/R$ in the film region.}}
\label{fig:wallpressure}
\end{figure}

{We plot in Fig. \ref{fig:wallpressure} the instantaneous liquid pressure profile on the drum wall for each of the cases studied here using \textit{Basilisk} at different timesteps, as the liquid film is entrained over it. Note that the x-axis is the curvilinear coordinate $s$, as defined previously in  Fig. \ref{fig:rotaryLLD_A1}(a), and the contribution from the atmospheric pressure $P_{atm}$ has been subtracted out, \i.e, $P_{w} = 0$ in air. The thick continuous line (black) represents wall pressure if the wheel were at rest. And so, it simply corresponds to the hydrostatic pressure underneath the liquid surface (for all $s < 0$) and the atmospheric pressure when $s \geq 0$ . As time increases, we observe that the pressure profiles very slightly deviate from the hydrostatic case, as long as $s \lesssim 0$. Thereafter, the wall pressure $P_w$ dips below the line $P_w = 0$ to reach a minimum, before raising back towards the atmospheric pressure. This further confirms the presence of an adverse pressure gradient ($G > 0$) between the meniscus and the film. In this overlap region, it can be remarked that the pressure profiles continue to evolve in time for all cases studied here. Insets in Fig. \ref{fig:wallpressure} display the snapshots of the VOF (red - liquid; blue - air; black - interface) at $t = 0.1$~s. As expected, the film thickness for the case when only density is quadrupled is much thinner than the other two cases since the thickness in the film region typically scales with the length $\delta_f \propto \sqrt{{\mu U}/{\rho \tilde{g}}}$, if the capillary number is $\mathcal{O}(1)$ \citep{JinAcrivosMunchPoF2005}. Moreover, in this zone, the centripetal acceleration on a fluid particle should be balanced out, at the leading order, by a radial pressure gradient \.i.e., $\partial p / \partial r = \rho u^2/R$. So, the pressure close to the wall should fall below $P_{atm}$ down to $\mathcal{O}\left(-\rho U^2 \delta_f/R\right)$, since the pressure at the free surface is equal to the atmospheric pressure due to the negligibly small contribution from interface curvature. This value is indicated by the horizontal dashed lines in Fig. \ref{fig:wallpressure}, which shows a good match with the pressure data from computations in a small region whose extent varies with time, $t$. As we move away from this zone, $P_w$ increases as we approach the front of the entrained liquid film and thereafter, $P_w = 0$ on the wall outside the film. {Finally, it is pointed out here that the maximum adverse pressure gradient evolves in time and attains a value $30\%$ smaller than what is estimated by the lubrication model (\ref{eqn:gradPavg_lubri_A1}). This might arise from the fact that (1) in simulations, the pressure profiles are very sensitive to the spatial resolution of the meniscus curvature and (2) in the model, the effect of inertia and the backflow are not included.}}

\begin{figure}
\begin{center}
\includegraphics[width=1\textwidth,keepaspectratio=true]{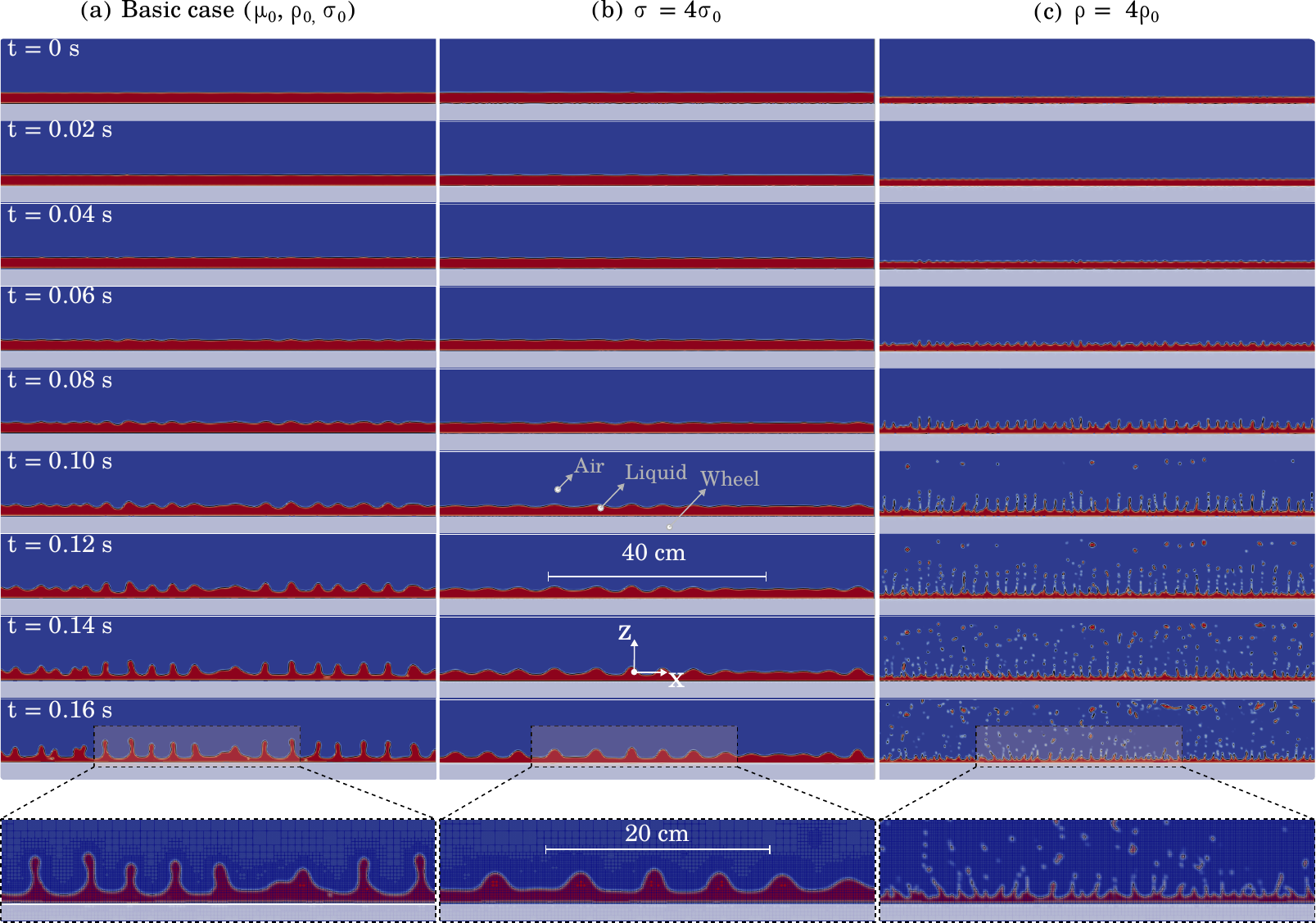}
\end{center}
\caption{Temporal evolution of the ribbing patterns in our direct numerical simulations using Basilisk \citep{popinet2009accurate, basiliskWebsite}. Here, the basic case corresponds to the water/UCON mixture ($\rho_{0} = 1044$kg~m$^{-3}$, $\mu_{0} = 100$Pa~s and $\sigma_{0} = 0.05$N~m$^{-1}$) at $U = 1.36$m~s$^{-1}$ and $H/R = 0.2$. The last row of images presents a zoom on the finger-like structures in order to illustrate the Adaptive Mesh Refinement for the last timestep $t = 0.16$s{. The maximum level of mesh refinement $N$ is equal to 9}. {}}
\label{fig:RibEvolution_Basilisk}
\end{figure}

\begin{figure}
\begin{center}
\includegraphics[width=1\textwidth,keepaspectratio=true]{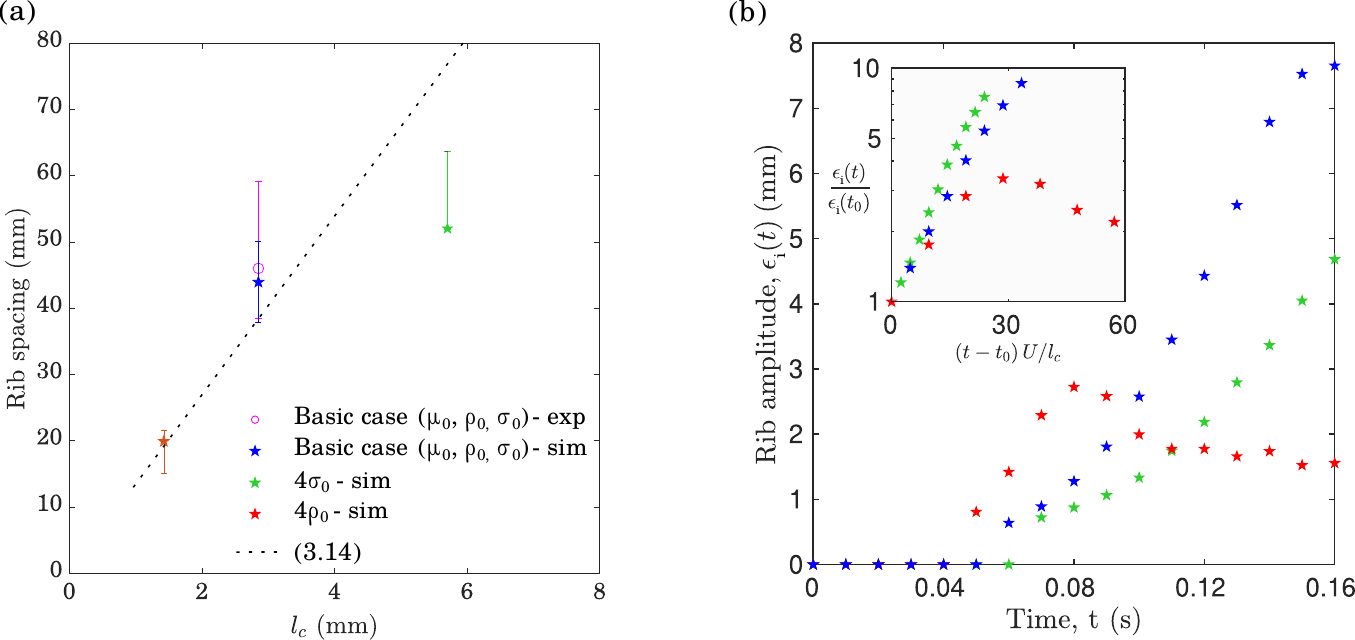}
\end{center}
\caption{(a) Comparison of rib spacing between {experiment ({\color{violet}$\medcircle$}),} simulations ($\medstar$) and the model (dotted line) \.i.e., $\lambda_*$ from (\ref{eqn:waveLen_ST}). {The experimental data corresponds to the one in Fig. \ref{fig:rib_spacing}(g): $U = 1.36$~m~s$^{-1}$ and $H/R = 0.2$ with $\rho_{0} = 1044$~kg~m$^{-3}$, $\mu_{0} = 100$~Pa~s and $\sigma_{0} = 0.05$~N~m$^{-1}$. } (b) Growth of interface corrugations as a function of time for the three cases considered in numerical simulations.}
\label{fig:RibSpacing_Basilisk}
\end{figure}

In order to provide more quantitative data {for the rib spacing}, the air/liquid interface positions are directly extracted from the VOF data at a fixed height from the bottom {($y = 67$~cm)} for various simulation times $t$. Fig. \ref{fig:RibEvolution_Basilisk} displays such spatio-temporal series of the VOF (blue: air and dark red: liquid), and the interface location (black continuous line). In all cases, the interface is initially flat and then evolves into an array of fingers at later times. These images from computations show a strong resemblance to the initial stage of digitation, or fingering patterns, when an initially flat interface between two immiscible viscous fluids in thin gaps is displaced by the least viscous fluid among the two. They also strikingly resemble the ribbing patterns in film-splitting flows \citep[see Fig. 9]{rabaud1994dynamiques}. {The last row of images in Fig. \ref{fig:RibEvolution_Basilisk} presents a zoom on the finger-like structures for the last timestep $t = 0.16$ in order to illustrate the Adaptive Mesh Refinement for the maximum level of mesh refinement $N$ equal to 9. The level of refinement is sufficient to capture the finest structures, whether in terms of amplitude or wavelength.} Similar to experimental measurements, the distance between ribs can be computed over the entire simulation time and the most probable rib spacing $\lambda$ is then deduced. Fig. \ref{fig:RibSpacing_Basilisk}(a) presents the comparison between such measured rib spacing $\lambda$ from computations and the ST rib spacing $\lambda_\ast$ (dotted line) as given by expression (\ref{eqn:waveLen_ST}), when the capillary number is $Ca \gtrsim 1$ ($\medstar$). The experimental rib spacing for the water/UCON mixture at $U = 1.36$~m~s$^{-1}$ and $H/R = 0.2$ is also given here ({\color{violet}$\medcircle$}). The error bars indicate the first and the last quartiles of the rib spacing distribution. {For the basic case, in addition to the maximum refinement level $N=9$, the level $N=10$ was also simulated. The value obtained for $\lambda$ with $N=10$ is within the error bar of that obtained with $N=9$.}

Firstly, the most probable rib spacing $\lambda$ from both the numerical simulation (basic case) and the experimental data match very well. So, the comparison between the model ($\lambda_\ast$) and measured spacing $\lambda$ for the basic case {is also favorable,} similar to what was already inferred for the experimental case, as seen for exemple in Fig. \ref{fig:rib_spacing}(c). 
{Secondly, in the other two cases considered in the numerical simulations, a good match with the model is seen for the case $\rho = 4\rho_0$, while the rib spacing is overestimated by the model when  $\sigma = 4\sigma_0$}. Finally, from the spatio-temporal VOF data in Fig. \ref{fig:RibEvolution_Basilisk}, it is also possible to recover the growth of corrugation amplitude on the air/liquid interface. Let $z_i(x, t)$ be the interface location at a given time.  Then, $\tilde{z}_i(x, t) = z_i(x, t) - \frac{1}{L_b}\int_0^{L_b} {z}_i(x, t) dx$  will represent the axial variation of the perturbation about a flat interface. Thereby, for each time, we define the perturbation amplitude $\epsilon_i(t)$ as the average along the z-axis of the root-mean-square such that $\epsilon_i(t) =  \sqrt{ \frac{1}{L_b}\int_0^{L_b} \tilde{z}^2_i(x, t) dx}$. The temporal evolution of such a perturbation amplitude is displayed in Fig. \ref{fig:RibSpacing_Basilisk}(b). For all cases, $\epsilon_i(t)$ is initially zero for a while, say up to some time $t_0$, and then grows very rapidly. Whereas the amplitude grows much faster for $\rho = 4\rho_0$ when compared with the basic case, it develops much slower for $\sigma = 4\sigma_0$. In addition, the interface corrugations $\epsilon_i(t)$ corresponding to the denser liquid case are observed to saturate. Indeed, as observed in Fig. \ref{fig:RibEvolution_Basilisk}(c), more and more tiny satellite droplets occur as the simulation advances from $t > 0.1$~s. This arises from insufficient spatial resolution in our simulations when $\rho = 4\rho_0$. We present in the inset of Fig.  \ref{fig:RibSpacing_Basilisk}(b), data points from all cases when the renormalized amplitude $\epsilon_i(t)/\epsilon(t_0)$ is plotted on a logarithmic scale against the non-dimensional time $\tau = U \left(t-t_0\right)/l_c$. The moderate data collapse on a linear trend implies that the renormalized amplitude increases exponentially in time and $l_c/U$ is the proper timescale for the development of ribbing patterns as in the case of the classical Saffman-Taylor instability. These observations further confirm the relevance of the proposed phenomenology for the ribbing pattern formation.


\section{Conclusions}
\label{sec:conclusions}
Using a wheel of radius $R = 19.5$~cm and $30$~cm width, we described axial flow patterns that appear along the wheel axis when it is rotated to drag-out liquid from a reservoir. Experiments were undertaken for two different liquid dynamic viscosities, namely, water and an approximately $100$ times more viscous water/UCON mixture. When the linear speed $U$ of the drum is increased systematically, the entrained coating flow over the drum presents rib-like patterns that consist of several liquid sheets stemming out of the bath on the rising side of the wheel. The same phenomenon occurs for various immersion depths of the wheel $H/R$. In all cases studied here, ribs emerge from the meniscus region in the neighbourhood where the liquid bath meets the wheel when the latter was at rest. Also, ribs often drift outward from the middle of wheel to the end walls. In general, we observed that the number of ribs approaches a \textit{normal} distribution while the distribution of distance between the sheets is skewed and presents a maximum at some characteristic length that we refer to as the most-probable rib spacing $\lambda$. Experimental measurements suggest that $\lambda$ does not vary much with the drag-out speed $U$ nor with the wheel immersion $H/R$ considered here. % The effect of the increase in liquid viscosity by about $100$-folds is to increase the observed rib spacing $\lambda$.

We also performed Direct Numerical Simulations using the \textit{open-source} software Basilisk \citep{popinet2009accurate, basiliskWebsite}, which solves the incompressible two-phase Navier-Stokes equations with an explicit interface capturing {method} (VOF method) for three case studies. {Computational domain is a cube of size $80$~cm, and a solid-body rotation of angular velocity $\Omega = 65$~rpm is applied on a half-cylinder of liquid to impose rotary drag-out. Simulations with the same liquid physical parameters as %{(le reference case est appelé basic case plus haut dans le document)}
the viscous water/UCON mixture on a $80$~cm long wheel, namely, the basic case, were first considered. The most-probable inter-rib distance obtained from computations showed excellent match with the experimental data. In the case where the only liquid density $\rho$ is modified to $4$ times the value in the basic case, the number of ribs is approximately doubled. In contrast, when only the liquid surface tension $\sigma$ is quadrupled in the simulations, the number of ribs decreased by a factor one-half.}

We considered a mechanism analogous to the directional Saffman-Taylor instability \citep{SaffmanTaylor1958, Hakim1990, Reinelt_DirectionalViscFingering_JFM1995} as the origin of such axial patterns. In the classical Saffman-Taylor instability, when some fluid drives out a more viscous fluid in thin gaps, the interface between both fluids exhibits finger-like protuberances that grow as the interface advances. In the present liquid drag-out problem, the {average} air/liquid meniscus is not displaced, but remains static. However, a fluid parcel in the neighbourhood between the film and meniscus region experiences an adverse pressure gradient since the liquid flow must accommodate local wall shear and mass conservation as it rises from the liquid bath towards the film region. Thereby, the meniscus region in the drag-out flow is prone to an instability analogous to the Saffman-Taylor instability (ST). We modelled the rib spacing via the critical ST wavelength $\lambda_{\ast}^2 = 12 \pi {\sigma}/ \vert G_\ast \vert$ wherein the driving pressure gradient $\vert G_\ast \vert$ was taken as the average adverse pressure gradient in the meniscus region of an equivalent inclined Landau-Levich-Deryaguin problem. Therefore, we proposed that the most probable rib spacing  be estimated as $\lambda_{\ast} = 2 \pi l_c \sqrt{3/\xi}$, where $\xi(Ca, H/R)$ is the non-dimensional {average} adverse pressure gradient for the corresponding inclined flat plate drag-out flow under lubrication approximation, and $l_c = \sqrt{\sigma/(\rho \tilde{g})}$ is the capillary length based on the effective gravity $\tilde{g}$, as computed from the plate inclination depending on the wheel immersion ratio $(H/R)$.

We demonstrated that the proposed rib spacing $\lambda_{\ast}$ is generally of the order of 15$l_c$, and it matches reasonably well with measurements in all our experiments, despite the inertial and complex two-phase nature of the problem, {where droplets and bubbles are formed between the liquid sheets}. Numerical simulations further confirmed the {proposed} scaling for the spacing and {for the growth rate of rib formation} at different liquid physical properties \i.e., the above-mentioned doubling and the halving of the most probable rib distance was well-captured by the capillary length scaling in $\lambda_{\ast}$. In summary, our analysis based on computations and experiments strongly confirms the ribbing pattern formation in rotary drum liquid entrainment is due to a mechanism analogous to the directional Saffman-Taylor instability \citep{Hakim1990, rabaud1994dynamiques, Reinelt_DirectionalViscFingering_JFM1995}. %We further illustrated that the most-probable inter-rib distance can be predicted from an estimate of the adverse pressure gradient via a lubrication model of an inclined drag-out flow.
{Our results also suggest that the most-probable inter-rib distance can be predicted from an estimate of the adverse pressure gradient via a lubrication model of an inclined drag-out flow. At this stage, it is needs to be further clarified why results from lubrication model can be relevant in such an inertial rotary LLD flow.} More work is necessary to highlight the role of the wheel radius, or more generally of external wall curvature, on the axial pattern formation during inertial liquid drag-out. Also, a full linear stability analysis for both the inclined plate LLD flow and the rotary drag-out problem, similar to that of \cite{Hosoi_RibRimming_1999PoF}, is needed to further validate and improve the proposed ST mechanism for ribbing patterns. %Also, kinematic waves leading to rivulet-like structures might also be an important factor at very small wheel immersion and low drum speeds, but they are quickly replaced by ribbing patterns. 
%{Comme pour les vidéos de manip, est-ce qu'on montre des vidéos obtenues avec Basilisk en supplementary material ?}

\textbf{Acknowledgements.}
Authors thank A. Buridon, G. Geniquet and S. Martinez from Universit\'{e} Claude-Bernard Lyon$1$ for their technical support in maintaining the experimental set-up and instrumentation. We also acknowledge intern students, namely, Rosie Cates (funded by LABEX iMUST), Élodie Duffaux and Solenne Gonin for their assistance with preliminary experiments and image processing drills.

\textbf{Funding.}
This work was partially supported by the LABEX iMUST of the University of Lyon (ANR-$10$-LABX-$0064$), created within the ``Plan France $2030$'' set up by the french government and managed by the French National Research Agency (ANR). The numerical part of this work was granted access to the HPC resources of PMCS$2$I (Pôle de Modélisation et de Calcul en Sciences de l'Ingénieur et de l'Information) of École Centrale de Lyon, and P$2$CHPD (Pôle de Compétence en Calcul Haute Performance Dédié) of Université Lyon $1$. 

\textbf{Declaration of interests.}
The authors report no conflict of interest.

\textbf{Author ORCIDs.}
J. John Soundar Jerome, \href{https://orcid.org/0000-0003-2148-9434}{0000-0003-2148-9434}; P. Trontin, \href{https://orcid.org/0000-0002-5191-2689}{0000-0002-5191-2689}; J.-P. Matas,\href{https://orcid.org/0000-0003-0708-1619}{0000-0003-0708-1619};

\bibliographystyle{plainnat}
\bibliography{SheetsLLD}
\end{document}